\documentclass{ametsocV61}
\usepackage{booktabs}
\usepackage{pdfpages}

\title{Knowledge-guided machine learning for disentangling Pacific sea surface temperature variability across timescales}

\authors{Kyle J. C. Hall\aff{a}\correspondingauthor{Kyle J. C. Hall, kylehall@umd.edu}, 
Maria J. Molina\aff{a}, 
Emily F. Wisinski\aff{a}, 
Gerald A. Meehl\aff{b},
and Antonietta Capotondi \aff{c,d}
}

\affiliation{\aff{a}{University of Maryland, College Park; College Park, MD}\\
\aff{b}{US NSF National Center for Atmospheric Research; Boulder, CO}\\
\aff{c}{Cooperative Institute for Research in the Environmental Sciences; University of Colorado Boulder; Boulder, CO}\\
\aff{d}{NOAA Earth System Research Laboratories; Boulder, CO}}

\abstract{Global weather and climate patterns are strongly influenced by dominant modes of anomalous Pacific sea surface temperature (SST) variability, including the El Niño-Southern Oscillation (ENSO), Pacific Meridional Mode (PMM), and Pacific Decadal Oscillation (PDO). However, disentangling these modes of variability remains challenging due to their spatial overlap and nonlinear coupling, which violate the assumptions of traditional linear methods. We develop a Knowledge-Guided AutoEncoder (KGAE) that uses spatiotemporal constraints and a gradient-based sparsity incentive to identify physically interpretable modes of detrended SST variability, each defined by a single broad characteristic timescale, without the need for predefined temporal filters or thresholds. The KGAE separates ENSO-like modes on 2- and 3–7-year timescales, as well as a decadal mode with characteristics reminiscent of the PDO and PMM, each with distinct spatial patterns. We perturb each latent dimension and use finite differences to characterize the state-independent and state-dependent sensitivities. We demonstrate that the decadal mode modulates ENSO diversity (central versus eastern Pacific) through both interference and nonlinear state dependence, and that a quasibiennial mode interacts with the interannual mode to characterize ENSO onset and decay. We assess the robustness of the KGAEs to random initialization and training stochasticity, to sampling bias induced by a time-stratified cross-validation scheme, and to unseen data (generalization). When applied to climate model output, KGAEs reveal model-specific biases in ENSO diversity and seasonal timing. Our results highlight how machine learning can uncover physically meaningful modes of Earth system variability and characterize their complex interactions across models and timescales.}

\begin{document}

\maketitle

%
%
%
%

%


\section{Introduction}\label{sec1}

Disentangling interannual and decadal Pacific sea surface temperature (SST) variability is challenging due to the overlapping spatial and temporal scales of ocean-atmosphere processes, yet it is crucial in order to be able to understand how they interact, to enhance predictive skill, and to protect human lives and economic prosperity \citep{Bjerknes_teleconnections, ropelewski_enso_teleconnections, Wei_PDO_Precip}. Advances in machine learning present an opportunity to revisit this longstanding challenge in characterizing SST variability \citep{molina2023review, eyring2024pushing}. We introduce a knowledge-guided machine learning framework to identify prominent modes of basin-wide variability at specific timescales and examine their nonlinear interactions. Our method identifies a mode of decadal variability, which is comparable in pattern and time evolution to the Interdecadal Pacific Oscillation \citep[IPO;][]{power1999inter} and Pacific Decadal Oscillation (PDO), as well as a set of two primarily tropical modes related to the El Niño-Southern Oscillation (ENSO) occurring at interannual and quasibiennial timescales. While the independence of North Pacific decadal variability and tropical interannual variability has long been debated \citep{di_lorenzo_modes_2023, Bjerknes_atmospheric_response, McPhaden_ENSO_on_top_of_background, wyrtki_ninodynamic_1975}, our method frames the problem as one of disentanglement rather than statistical independence, and enables us to directly characterize cross-timescale nonlinear modulations which linear methods, by construction, do not address.

Early studies postulated that there could be a fundamental divide between extratropical and tropical Pacific variability, with the former operating primarily on decadal time scales and the latter on interannual time scales \citep{yeh_tropical_extratropical_sst}. However, more recent work has noted that coupled tropical-extratropical interactions occur across nearly all timescales \citep{power2021decadal, meehl2006megadroughts, zhao_removing_2021}. This emerging understanding motivates the development of methods to identify basin-wide signatures of the leading modes of Pacific variability and quantify their interactions. ENSO, for example, is often framed as an interannual (2-7 year) tropical phenomenon, yet its indices exhibit decadal variability and links to extratropical processes \citep{Zhang_Ensolike_decadal_var, Capotondi_TPDV, Deser_interdecadal_tropics_northpac_linkage}. Conversely, the PDO is usually defined as a decadal extratropical pattern (7+ years), but it is closely associated with the IPO, which spans both tropical and midlatitude variability, a signal reflected in PDO indices \citep{power1999inter, Alexander_AtmosphericBridge, Newman_PDO_revisited}. This spectral and spatial overlap complicates any view of ENSO and PDO as distinct, cleanly separable phenomena. At the same time, although many studies recognize the interdependence of these modes \citep[e.g.,][]{chang_pacific_2007, Vimont_interannual_impact_on_decadal, Deser_ENSO_PDV_CESM, meehl2021role}, other work suggests that the IPO has mechanisms distinct from ENSO, even if ENSO events may help trigger IPO transitions \citep{meehl2021role}. These findings, together with entrenched conventions in the literature, have encouraged a view of Pacific SST variability as a set of separate `modes,' often at the expense of a more holistic characterization. Accordingly, we argue that methods based on assumptions of linearity, orthogonality, and variance maximization, or on a priori geographical regions or frequency bands, provide only a partial view of Pacific variability \citep{deser_teleconnectivity_2000, monahan_eofsbad_2009, dommenget_pca_caution_2002, fulton_pca_nonphysical_2021}. 

Previous research has sought to address these limitations. Rotated EOF methods, including varimax rotation, can reduce spatial or temporal complexity and improve interpretability \citep{richman_rotation_1986, mestas-nunez_orthogonality_2000, Kaiser_1958, lian_evaluation_2012}. Extended EOFs incorporate time lags to better represent evolving structures \citep{weare_examples_1982}. Other approaches, such as singular spectrum analysis, maximum covariance analysis, and pairwise rotations, relax different aspects of standard EOF analysis and can help separate patterns that are otherwise mixed in space, time, or across fields \citep{chen_pairwise-rotated_2017}. Some studies have tailored EOF-based methods to specific scientific goals by rotating and recombining EOFs with regression-based approaches. For instance, \citet{takahashi_e_and_c_patterns} explored ENSO diversity, while \citet{chen_orthogonal_2016} derived orthogonal ENSO and PDO indices that maximize different statistical properties through regression. Low Frequency Component Analysis identifies low-frequency patterns by linearly recombining EOFs to maximize the ratio of low-frequency variance to total variance \citep{wills_disentangling_2018,wills_ocean_2019}. Linear Inverse Model (LIM) approaches identify dynamical eigenmodes that are not constrained to be orthogonal, but the overlap among eigenmodes does not account for possible nonlinearities \citep{penland_studies_2006}. Bayesian methods have also been used to extract nonlinear principal manifolds from SST data \citep{mukhin_principal_2015}. These methods have yielded important insights, but still rely on assumptions that may constrain how faithfully they represent the underlying climate dynamics.

To overcome these limitations, we propose a Knowledge-Guided AutoEncoder (KGAE), an unsupervised generative neural network that relaxes assumptions of orthogonality and linearity while incentivizing the spatiotemporal disentanglement of captured variability. Unlike conventional linear decompositions, which typically impose orthogonality and favor uncorrelated, statistically independent, variance-maximizing modes, the KGAE separates spatial and temporal criteria when disentangling learned modes of variability, allowing those modes to interact through nonlinear state dependence. In this framework, temporal dependence is defined as overlap in the power spectra of two modes of variability (i.e., two modes oscillating on shared frequencies), while spatial dependence is defined as correlation between their associated spatial patterns. Temporal independence is prioritized over spatial independence, with the penalty on spatial dependence applied only when temporal dependence is nonzero (Appendix A). By adopting this relaxed and decoupled formulation of ``spatiotemporal disentanglement," the KGAE learns physically meaningful modes of Pacific SST variability associated with broad characteristic timescales (e.g., interannual versus decadal) and captures the ways they interact without relying on prescribed filters or thresholds.

\section{Methods}\label{sec3}

\subsection{$\beta$-Variational Autoencoder Formulation and Network Architecture}

\begin{figure}[h]
\centering
\includegraphics[width=0.95\linewidth]{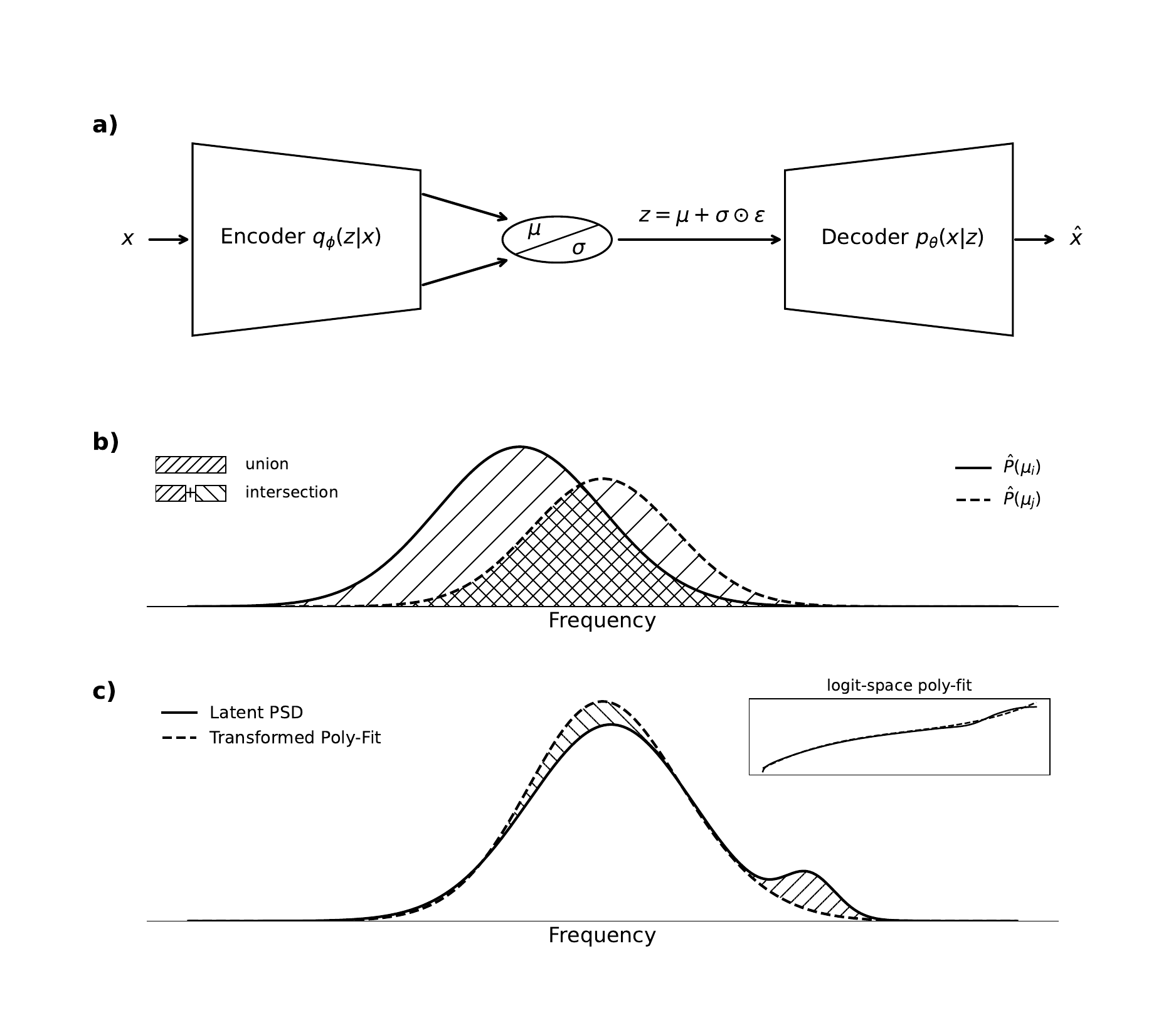}
\caption{Schematics of the: a) $\beta$-VAE architecture used by KGAEs, with all modifications implemented through the loss function; b) spectral overlap loss, defined as the intersection-to-union ratio of two latent power spectra; c) spectral modality loss, defined from the residual between the normalized latent power spectrum and a reconstructed unimodal spectrum derived from a fourth-degree polynomial fit to the logit-transformed cumulative power spectrum.}
\label{fig:1}
\end{figure}

KGAEs are based on the $\beta$-variational autoencoder \citep[$\beta$-VAE;][]{diederik_introduction_2019, kingma2022autoencodingvariationalbayes, higgins2017betavae}. Given an input $\mathbf{x}$, the encoder maps $\mathbf{x}$ to a distribution over latent representations $\mathbf{z}$, and then the decoder maps samples from this distribution to reconstructions $\hat{\mathbf{x}}$. Specifically, they parameterize (i) an approximate posterior distribution over latent dimensions $\mathbf{z}$, 

\begin{equation}
    q_\phi(\mathbf{z}\mid \mathbf{x}) = \mathcal{N}\left(\boldsymbol{\mu}(\mathbf{x}), \operatorname{diag}\left(\boldsymbol{\sigma}^2(\mathbf{x})\right)\right),
\end{equation}

\noindent where parameters $\phi$ define the encoder network, and (ii) a conditional likelihood $p_\theta(\mathbf{x}|\mathbf{z})$, where the parameters $\theta$ define the decoder network. In general, $\mathbf{z}$ has substantially lower dimensionality than $\mathbf{x}$, thereby serving as an information bottleneck (Figure \ref{fig:1}a). 

A standard VAE maximizes the evidence lower bound (ELBO) on the marginal log-likelihood $\log p_\theta(\mathbf{x})$. The $\beta$-VAE modifies this objective by weighting the Kullback--Leibler (KL) divergence:

\begin{equation}
\mathcal{L}_{\beta}(\theta, \phi; \mathbf{x}) 
= \mathbb{E}_{q_\phi(\mathbf{z} \mid \mathbf{x})} \big[ \log p_\theta(\mathbf{x} \mid \mathbf{z}) \big] 
- \beta\, \mathrm{KL} \big( q_\phi(\mathbf{z} \mid \mathbf{x}) \, \| \, p(\mathbf{z}) \big),
\end{equation}

\noindent where $p(\mathbf{z})$ is an isotropic Gaussian prior, typically $\mathcal{N}(\mathbf{0},\mathbf{I})$, and $\mathrm{KL}(\cdot|\cdot)$ denotes the Kullback-Leibler divergence between the approximate posterior and the prior. In practice, training proceeds by computing encoder outputs $\boldsymbol{\mu}(\mathbf{x})$ and $\log \boldsymbol{\sigma}^2(\mathbf{x})$, and then sampling from $q_\phi(\mathbf{z}\mid \mathbf{x})$ using the `reparametrization trick,'

\begin{equation}
    \mathbf{z} = \boldsymbol{\mu}(\mathbf{x}) + \boldsymbol{\epsilon} \odot \boldsymbol{\sigma}(\mathbf{x}), \qquad \boldsymbol{\epsilon} \sim \mathcal{N}(\mathbf{0}, \mathbf{I}),
\end{equation}

\noindent where $\mathbf{I}$ is the $d\times d$ identity matrix and $d$ is the latent dimensionality. The sampled latent vector $\mathbf{z}$ is then passed through the decoder to obtain the reconstructed mean, $\hat{\mathbf{x}} = D_\theta(\mathbf{z})$. 

For a conventional $\beta$-VAE, training minimizes the negative $\beta$-VAE objective, comprising the negative reconstruction log-likelihood and the weighted KL divergence term. When $p_\theta(\mathbf{x}\mid \mathbf{z})$ is Gaussian with fixed variance, the negative reconstruction log-likelihood is proportional to the mean squared reconstruction error. The baseline $\beta$-VAE loss can therefore be written, subject to the corresponding scaling of its terms, as

\begin{equation}
    \mathcal{J}_{\beta}(\theta,\phi;\mathbf{x})
=\mathbb{E}_{q_\phi(\mathbf{z}\mid\mathbf{x})}\left[
\operatorname{MSE}(\hat{\mathbf{x}},\mathbf{x})
\right] + \beta\, \mathrm{KL} \big( q_\phi(\mathbf{z} \mid \mathbf{x}) \, \| \, p(\mathbf{z}) \big). 
\end{equation}

\noindent This baseline loss does not represent the complete KGAE training objective. KGAEs augment $\mathcal{J}_{\beta}$ with the knowledge-guided and regularization terms introduced in the following subsection to encourage spatiotemporal disentanglement.

For this work, the encoder consists of two fully connected hidden layers with 248 nodes each and hyperbolic tangent activation functions, followed by a 5-dimensional linear latent representation, with each latent dimension parameterized by a $\mu$ and $ln(\sigma^2)$ (Fig. \ref{fig:1}a). The decoder comprises two fully connected hidden layers with 248 nodes each and hyperbolic tangent activation functions, followed by a linear output layer with one node per grid point (13,739 total in the Pacific basin). 

\subsection{Knowledge-Guided Loss Function Terms}

To examine the interdependencies between decadal and interannual Pacific SST variability, we developed a set of additional loss terms designed to encourage specific statistical and spectral structure in the latent space. Although definitions of independence and disentanglement vary, our guiding assumption is that patterns of variability sharing similar spatial patterns or spectral characteristics are more likely to reflect related underlying dynamics than variability with distinct spatial patterns and timescales. Motivated by this idea, we designed the loss so that each latent dimension preferentially captures a single mode of variability characterized by a broad timescale (e.g., interannual or decadal), together with an associated spatial pattern, without requiring a priori specification of frequency bands or geographic regions. 

In addition to the baseline $\beta$-VAE objective, comprised of reconstruction and Kullback-Leibler divergence terms, we first encourage unimodality in each latent dimension's power spectrum by penalizing the residual between the normalized spectrum and a reconstructed unimodal spectrum derived from a fourth-degree polynomial fit to the logit-transformed cumulative power spectrum (Fig. \ref{fig:1}c). This term encourages each latent dimension to correspond to a single characteristic timescale. We then penalize the spectral overlap and spatial correlation among latent dimensions. Since the Fourier transform is a differentiable linear operator, we can compute and penalize the intersection-over-union of each pair of latent power spectra (Fig. \ref{fig:1}b). For pairs of latent dimensions with overlapping spectra, we additionally penalize correlation between their associated spatial patterns. These spectral-overlap and conditional spatial-correlation penalties are weighted by ``salient variance,'' which we define as the normalized magnitude of the gradient of the network output variance with respect to each latent dimension. Following the intuition of sparse autoencoders \citep{cunningham2023sparseautoencodershighlyinterpretable}, which promote selective use of higher-dimensional representations through sparsity, this weighting allows the KGAEs to down-weight latent dimensions that contribute relatively little to reconstruction. 

During development, we tested several configurations of these loss terms, including correlation alone, spectral overlap alone, and both together. Penalizing spatial correlation alone yielded modes resembling EOF-like decompositions, whereas penalizing spectral overlap alone yielded multiple ENSO-like modes, including decadal ENSO-like variability, akin to applying PCA to band-filtered data. We found that neither constraint alone yielded modes defined primarily by their characteristic timescale, which was our goal. Instead, variability associated with dominant modes (characterized primarily by other timescales but displaying cross-spectrum variance) leaked into the other modes. Without these additional constraints, the latent dimensions are not uniquely identifiable, and post hoc transformations of the latent space would be needed, complicating interpretation. 

The spectral modality and spatiotemporal disentanglement loss terms are applied to the latent posterior mean $\boldsymbol{\mu}$ rather than the sampled latent variable, $\boldsymbol{z} = \boldsymbol{\mu} + \boldsymbol{\sigma}\odot\boldsymbol{\epsilon}$, because $\boldsymbol{\epsilon}$ is defined in the variational framework as i.i.d. Gaussian noise. Applying cross-sample constraints directly to the sampled latent variable would therefore act on stochastic noise realizations rather than on the learned latent structure itself. Although sampled latent vectors $\boldsymbol{z}$ are used during training, all subsequent analyses use the latent posterior mean $\boldsymbol{\mu}$ as a deterministic representation. This is equivalent, for evaluation, to setting $\boldsymbol{\epsilon}=0$ in the reparameterization equation, such that $\boldsymbol{z} = \boldsymbol{\mu}$. Finally, we include a loss term that penalizes the magnitude of the first-layer decoder weights and another that penalizes deviations of the (mini-batch mean) latent posterior mean from zero. Together with the KL-divergence term, these terms serve as regularization and improve KGAE training stability. More details are available in Appendix A. 

\subsection{Data and Problem Design}

We conducted two primary experiments to evaluate the robustness of the KGAEs against different sources of uncertainty and investigate Pacific SST anomaly (SSTA) variability across timescales. The first experiment, hereafter the ``large ensemble" experiment, identifies basin-wide Pacific SSTA signatures of decadal, interannual, and quasibiennial modes and characterizes their nonlinear state-dependence using latent basis-vector finite-difference analysis. This experiment leverages a large ensemble ($n=100$) of randomly initialized KGAEs to assess the sensitivity of the learned patterns to random weight initialization and stochasticity during training, and uses a time-stratified cross-validation scheme to evaluate robustness to data record sub-sampling. The large-ensemble KGAEs were trained, validated, and tested on detrended monthly SSTAs from the fifth-generation ECMWF atmospheric reanalysis (ERA5) from January 1940 through December 2023 \citep{hersbach_era5_2020, soci2024era5, hirahara201626}, which leverages HadISST 2.1.0.0 \citep{hadsst} before September 2007, and the Met Office OSTIA dataset thereafter \citep{donlon_operational_2012}. The sub-periods used for training, validation, and testing are discussed in the next subsection.

Two separate bootstrapping protocols were used to assess different sources of result uncertainty. For analyses assessing decoder sensitivities and the temporal characteristics of the modes learned by the KGAEs, we present the ensemble-mean decoder sensitivities and latent time series. In those analyses, we use the ensemble spread and confidence intervals derived from i.i.d. bootstrap replicates ($n=1000$) of the ensemble, sampling members with replacement, to quantify result uncertainty arising from random initialization and training stochasticity. For analyses assessing the strength of the signals learned by the KGAEs in ERA5, we instead use ensembles of 12-month temporal block-bootstrap replicates ($n=1000$) to quantify result uncertainty arising from finite sampling. We use 12-month blocks to offset biases that may arise from high short-lag temporal auto-correlations (i.e., the seasonal persistence of ENSO). The bootstrapping method used in each analysis is denoted in the respective figure captions. 

The second experiment, hereafter the ``progressive split'' experiment, uses nine progressively expanding contiguous train/test splits to quantify out-of-period generalization. Across these splits, the training fraction increases from 50\% to 90\% in increments of 5\%, while the held-out test fraction correspondingly decreases from 50\% to 10\%, with five random initializations used for each model configuration ($n=5$). The experimental procedure and results from the progressive-split experiment are described in Section 3\ref{exp2resultshere}. All other results use KGAEs from the large-ensemble experiment.

\subsection{Interleaved Time-Stratified Cross Validation}

For the large-ensemble experiment, we used January 1940 to December 2014 as the combined training and validation period and January 2015–December 2023 as the held-out test period. Within the training and validation period, KGAEs were trained with an interleaved, time-stratified variant of five-fold cross-validation. Rather than dividing the record into five contiguous blocks, we assigned every fifth monthly sample to the same non-overlapping fold (Supplemental Figure 1). Each fold therefore spans the full record with a regular five-month sampling interval, thereby preserving access to low-frequency variability needed for the Fast Fourier Transform (FFT) and spectral analysis. In contrast, standard five-fold cross-validation with contiguous folds would produce fold lengths of about 15 years, limiting the ability of the corresponding power spectra to resolve decadal variability. This interleaved, time-stratified cross-validation design also reduces seasonality-related sampling bias, since each fold contains every fifth month over the full record, so each training set repeatedly samples all calendar months over time. 

This approach does, however, introduce tradeoffs. Since all folds span the same period, short-lag autocorrelation and shared low-frequency variability create dependence between the training and validation subsets. However, this dependence is less problematic than it would be in a predictive setting because our goal is robust characterization of variability, not out-of-sample prediction. Additionally, the five-month interval between samples may allow a given fold to ``skip" an extreme event captured by other folds. For example, if a fold samples October prior to a strong ENSO event, it may not sample the event peak during the following NDJF season. As a result, fold-specific patterns may be less sensitive to extreme events. We assess this effect by comparing per-fold spatial patterns from the large ensemble with corresponding patterns obtained from models trained on the full period. 

During each iteration of interleaved, time-stratified five-fold cross-validation, one fold is held out for validation, and the remaining four are used for training. Trend and climatology are estimated exclusively from the four training folds and then removed from the corresponding training, validation, and test sets (see next subsection). Each fitted KGAE was used to encode and reconstruct its validation fold and the held-out 2015–2023 test period. Although test data are not typically concatenated with validation output in predictive machine-learning workflows, this step was necessary here to extend the latent time series sufficiently to assess decadal variability. We also trained a sixth KGAE on the full 1940-2014 period as a baseline for comparison to the cross-validated models.

\subsection{Preprocessing and Training}

For each interleaved, time-stratified cross-validation KGAE, the trend and monthly climatology estimated from the four training folds are removed from the training, validation, and test sets. To estimate the trend, we fit a second-degree polynomial to the latitudinally weighted basin-averaged SSTA of the training set and subtract the resulting time series from each grid cell in the respective training, validation, and test sets. Although the nonstationarity of climate variability is an active area of research \citep{xing_et_all, cluett_pan-basin_2025, klavans_human_2025}, we detrend to more clearly reveal variability on shorter timescales. The basin-average trend was used to preserve the spatial coherence of SSTA patterns. Results were not sensitive to whether basin-wide or grid cell-specific trends were used, nor to whether detrending occurred before or after deriving anomalies. No further scaling was applied to preserve the units of the SSTAs (Kelvin), which facilitated the interpretation of the results. For the full-period baseline model, preprocessing quantities are estimated from the complete 1940–2014 training period.

To assess sensitivity to initialization, during the large-ensemble experiment, weights were randomly initialized 100 times for each of the five time-stratified cross-validation iterations using Xavier Normal weight initialization \citep{pmlr-v9-glorot10a}, and biases were initialized to zero. For the progressive-split experiment, we trained five randomly initialized KGAEs for each of the nine contiguous train/test splits using the same procedure. All KGAEs were trained for 200 epochs using exponential moving averages of the network weights with a decay factor of 0.995 \citep{moralesbrotons2024exponentialmovingaverageweights}. Exponential moving averages of the network weights were used for validation during training and for post-hoc evaluation. Since the learned KGAE modes are not ordered or signed a priori, they must be aligned across folds and random initializations, analogous to the sign alignment commonly applied to EOFs. We therefore aligned the sign of each latent mode according to the sign of its spatial regression onto the Niño-3.4 index, using spatial patterns defined by the state-independent decoder sensitivities (discussed in Section 2\ref{decsensitive}), and then sorted the modes by characteristic timescale.

\subsection{Hyperparameter Tuning}

We performed several rounds of hyperparameter tuning to identify a KGAE architecture that best separates the timescales of Pacific SSTA variability. Model selection prioritized low spectral overlap, low reconstruction loss, robustness to random weight initializations, and stability during interleaved, time-stratified cross-validation. We began with a broad, random, and manually guided grid search over activation functions, number of hidden layers, latent dimensionality, and hidden layer width. In total, tens of thousands of cross-validated model fits were used to narrow the hyperparameter space. We then performed several single-objective optimization experiments using Bayesian optimization with the Tree-structured Parzen Estimator algorithm and our custom loss as the objective. We also conducted several multi-objective optimization experiments using the Nondominated Sorting Genetic Algorithm II and various combinations of the terms in our custom loss. The architecture selected through this process was used in the final analysis. 

\subsection{Decoder Sensitivity Analysis}\label{decsensitive}

We use latent finite differences along individual latent dimensions to assess how the trained KGAE decoder responds to each latent mode. This approach is based on latent traversals, which isolate the effect of individual latent dimensions on the decoder output \citep{higgins2017betavae}. Similar finite-difference methods have been used to distinguish symmetric and asymmetric responses to climate indices \citep{hoerling_nonlinear_composites, monahan_spatial_2004, deser_northern_2017, jimenez-esteve_nonlinearity_2019}. Since we center the analysis around the origin, equal-and-opposite responses to positive and negative perturbations are interpreted as the linear-response component, whereas departures from this symmetry provide a measure of nonlinearity.

For each latent dimension $\mathbf{z}_j$, we define a vector $\mathbf{e}_j$ that equals 1 in the $j$th dimension and 0 in all other dimensions. We then decode positive and negative perturbations along this dimension. Let $D(\mathbf{e}_j)$, $D(-\mathbf{e}_j)$, and $D(\mathbf{0})$ denote the decoder outputs for the positive perturbation, negative perturbation, and origin, respectively. We center this analysis at the origin because the KGAE training objective includes an explicit centering penalty (Equation A22) and regularization toward an isotropic Gaussian prior (Equation 4). Consistent with these constraints, the latent time series have means near zero (approximately 0.01) and standard deviations near one (approximately 0.9). 

We separate the decoder response into symmetric and asymmetric components. The symmetric component, which we refer to as the linear response, is

\begin{equation}
    \mathrm{L}_j=\frac{D(\mathbf{e}_j) - D(0)}{2} + \frac{D(0) -D(-\mathbf{e}_j)}{2} = \frac{D(\mathbf{e}_j)-D(-\mathbf{e}_j)}{2}
\end{equation}

The asymmetric component, which measures the nonlinear response, is

\begin{equation}
    \mathrm{NL}_j=\frac{D(\mathbf{e}_j) -D(0)}{2} - \frac{D(0) -D(-\mathbf{e}_j)}{2} =\frac{D(\mathbf{e}_j) + D(-\mathbf{e}_j) - 2D(0)}{2}
    \label{nlreqn}
\end{equation}

To assess how the decoder response to mode $\mathbf{z}_j$ depends on the state of another mode $\mathbf{z}_{j'}$, we calculate the nonlinear response to mode $\mathbf{z}_j$ while holding $z_{j'}$ at $+1$ and $-1$. Their half-difference defines the state-dependent modulation:

\begin{equation}
    \text{State-Dependent Modulation}_{j \leftarrow j'} = \frac{(NL_j | z_{j'}=1) - (NL_j | z_{j'}=-1)}{2}. \label{statedependenteqn}
\end{equation}

\noindent We also assess whether the decoder response to a mode depends on the state of that same mode. For this analysis, we calculate local nonlinear responses centered at $z_j=-1$ (NLNR) and $z_j=1$ (PLNR) using perturbations of magnitude 0.25:

\begin{equation}
    \mathrm{NLNR}_j=\frac{D(-1.25\mathbf{e}_j) + D(-0.75\mathbf{e}_j) - 2D(-\mathbf{e}_j)}{0.5},
\end{equation}

\begin{equation}
    \mathrm{PLNR}_j=\frac{D(1.25\mathbf{e}_j) + D(0.75\mathbf{e}_j) - 2D(\mathbf{e}_j)}{0.5},
\end{equation}

\begin{equation}
    \text{Self-State Dependence}_j = \frac{\mathrm{PLNR}_j - \mathrm{NLNR}_j}{2}.
    \label{selfstatedependenteqn}
\end{equation}

\noindent The state-dependence modulation in Equation (\ref{statedependenteqn}) measures how the state of one mode alters the decoder response to another mode. In contrast, the self-state dependence in Equation (\ref{selfstatedependenteqn}) measures how the decoder response to a mode varies between the negative and positive portions of its own latent axis. Thus, the overall nonlinear response in Equation (\ref{nlreqn}) can be small when the negative and positive local responses have opposite signs, even though the decoder exhibits strong self-state dependence. These quantities characterize the local finite-difference behavior of the decoder and do not represent temporal evolution. 

\subsection{Quantitative Comparison of Spatial Patterns}

As noted above, KGAE nonlinearity allows the spatial pattern associated with a given latent dimension to vary with both its own value and the values of the other latent dimensions. Accordingly, the relationship between a KGAE latent mode and a conventional climate mode cannot be characterized solely by the temporal correlation between their associated time series. Instead, we compare their spatial patterns. We quantify the similarity between two spatial patterns using their latitude-weighted spatial Pearson correlation, computed across grid points. 

We compare spatial patterns derived from the KGAE experiments with spatial patterns associated with several commonly cited modes of Pacific climate variability. These reference modes include three ENSO indices, namely E- and C-Indices of \citet{takahashi_e_and_c_patterns} and Niño-3.4 \citep{trenberth_definition_of_el_nino_1997}; the Pacific Decadal Oscillation (PDO) index \citep{Mantua_PDO}; the North Pacific Gyre Oscillation (NPGO) index \citep{di_lorenzo_north_2008}; and the Pacific Meridional Mode (PMM) index \citep{chiang_analogous_2004}. Where possible, we computed these indices directly from the detrended ERA5 SSTA training data following the cited definitions. For the NPGO and PMM, which are not defined from SST alone, indices were obtained from the NOAA Physical Sciences Laboratory. 

For each reference mode, we obtained a basin-wide SSTA pattern by regressing the detrended ERA5 SSTA training data onto the corresponding index (Supplemental Figure 2). For the PMM, we first removed SSTA variability associated with the Cold Tongue Index (CTI), following  \citet{chiang_analogous_2004}. Since both the reference modes and the KGAE modes vary in their regional concentration, we evaluate spatial similarity over three domains (Table \ref{tab:kgae_climate_panels}): the full Pacific basin, tropics (23$^{\circ}$S-23$^{\circ}$N), and North Pacific ($\geq21^{\circ}$N). 

\section{Results} 

\subsection{Basin-wide Pacific SST Variability in a Cross-Validated Large Ensemble}

\begin{figure}[h]
\centering
\includegraphics[width=0.75\linewidth]{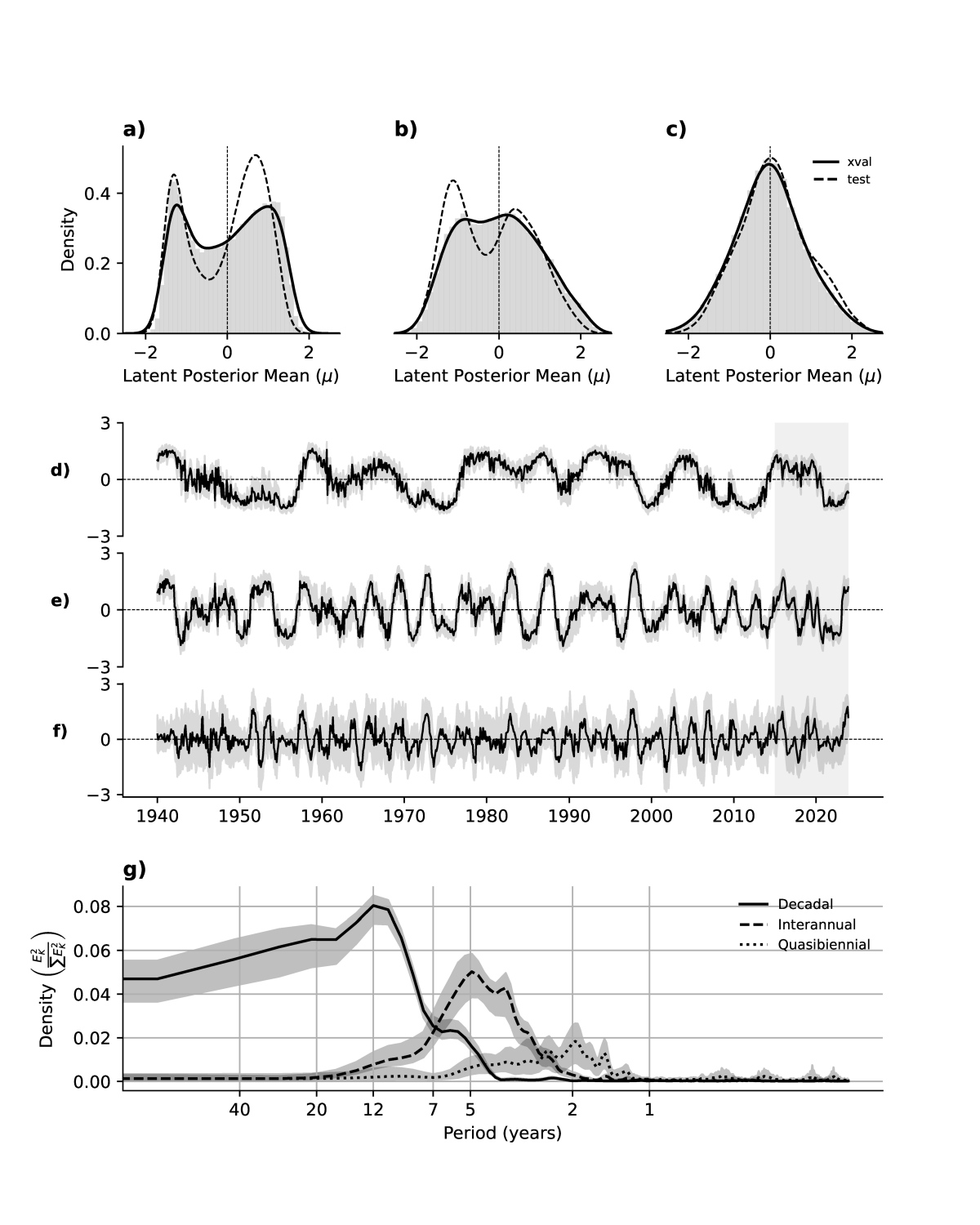}
\caption{a-c) Distributions of latent posterior means for a) decadal, b) interannual, and c) quasibiennial modes. Gray bars show density-normalized histograms of the cross-validation (xval) period. Solid and dashed black curves show Gaussian kernel density estimates for the xval and test periods, respectively. d-f) Time series of the latent posterior mean $\mu$ corresponding to the KGAE d) decadal, e) interannual, and f) quasibiennial modes. The light-gray shaded area at the end of the time series denotes the held-out test period. g) Power spectral density curves for KGAE modes as indicated in the legend. The gray shading in d-g) indicates the two-tailed 95\% confidence level from the KGAE ensemble members ($n=100$); no bootstrapping was used.}
\label{fig:2}
\end{figure}

\begin{figure}[h]
\centering
\includegraphics[width=1\linewidth]{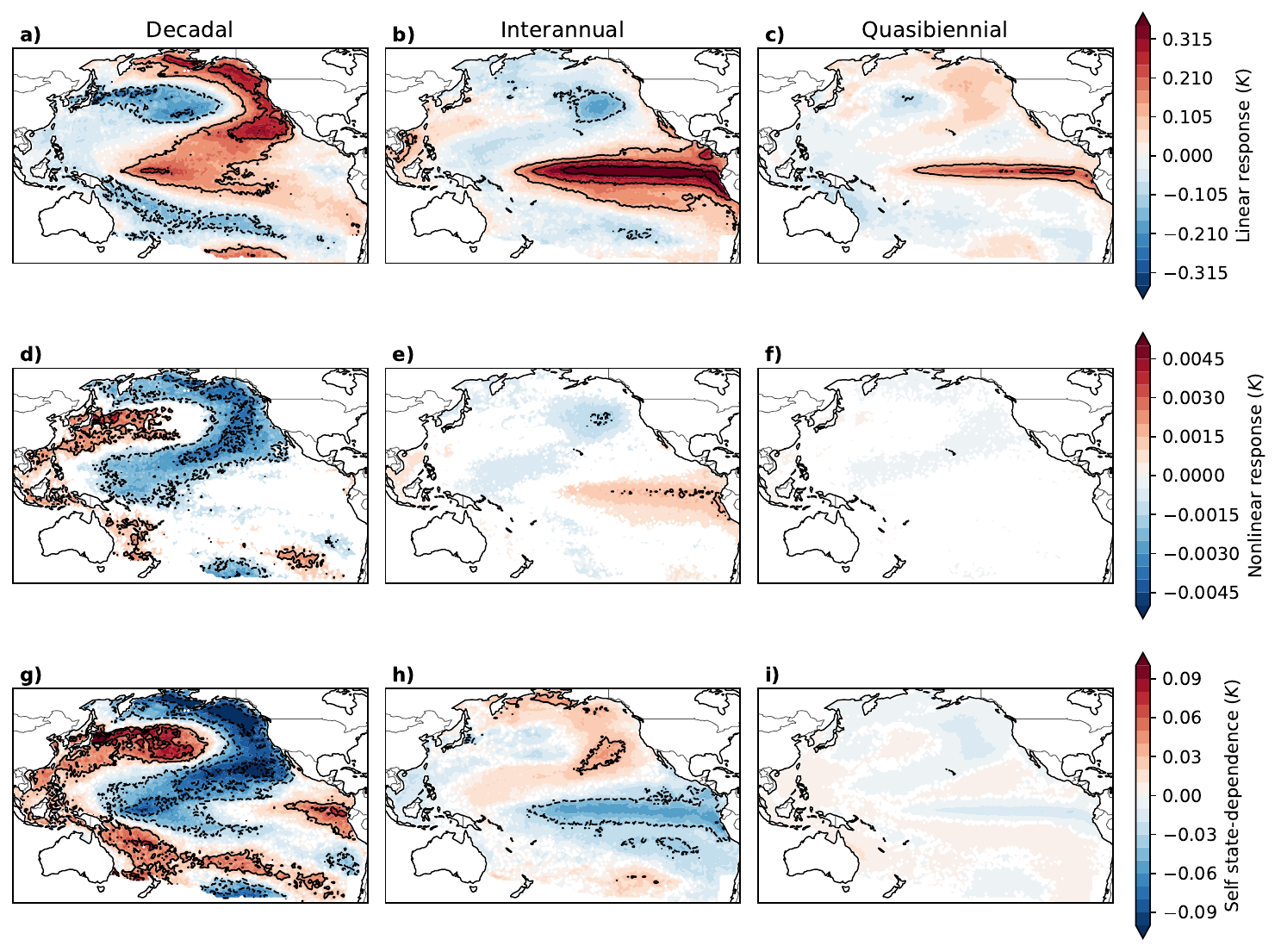}
\caption{Large ensemble mean ($n=100$) of the finite-difference responses of the KGAE decoder. a-c) Linear-response components for the decadal, interannual, and quasibiennial latent modes, respectively. d-f) Corresponding nonlinear responses. g-i) Self-state dependence of the decoder responses for the same three latent modes. Only values with 95\% confidence intervals that exclude zero are shown. Confidence intervals were calculated from 1,000 bootstrap replicates generated by independently resampling the KGAE large-ensemble members.}
\label{fig:3}
\end{figure}

\begin{figure}[h]
\centering
\includegraphics[width=1\linewidth]{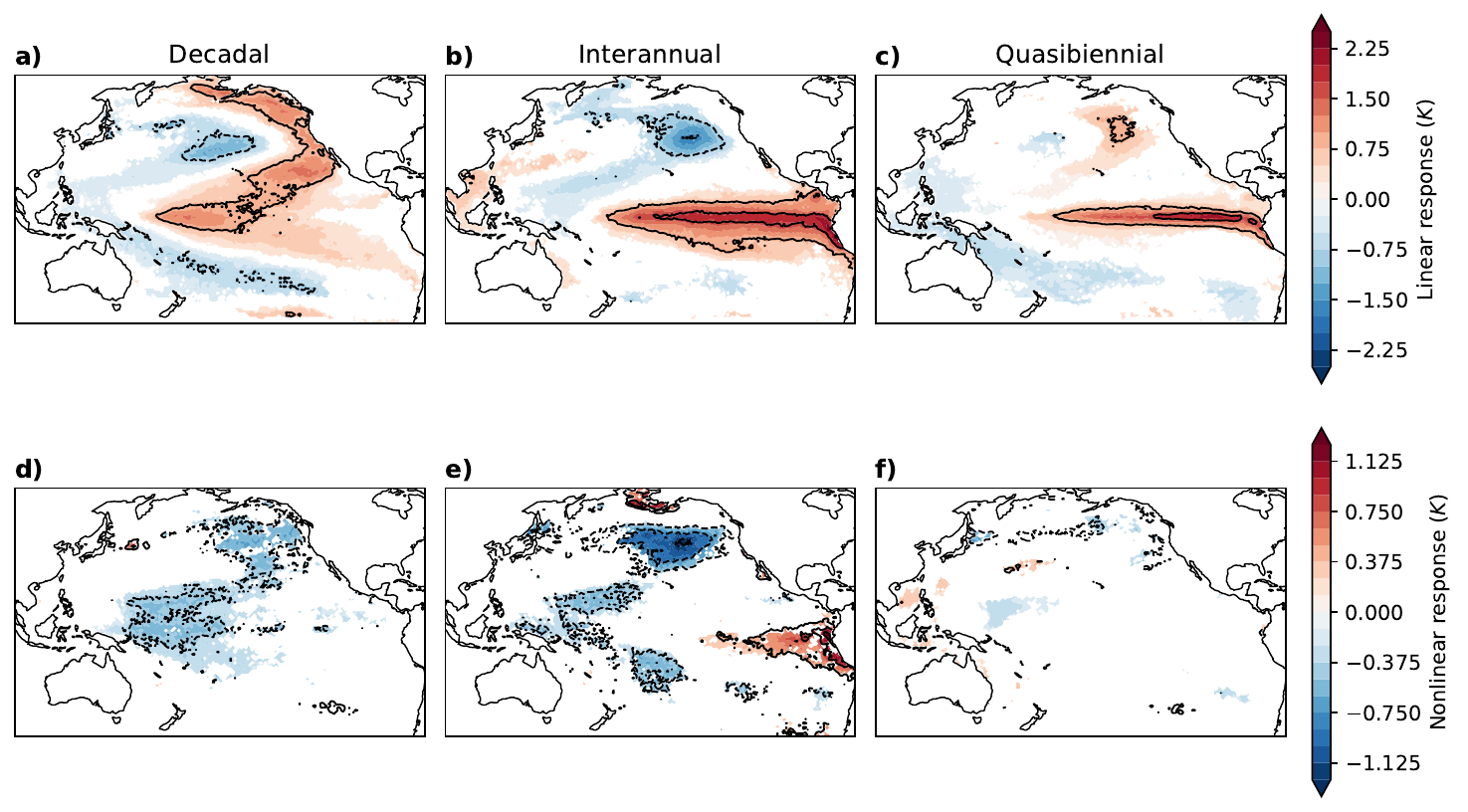}
\caption{Composite-based responses of detrended ERA5 SSTAs associated with the ensemble-mean ($n=100$) KGAE modes. a-c) Linear-response components for the decadal, interannual, and quasibiennial latent modes, respectively, computed from their positive and negative phases ($|z|>1\sigma$) while holding other modes neutral ($|z|<1\sigma$). d-f) Corresponding nonlinear responses, computed from the positive, neutral, and negative phases delineated by $\pm1\sigma$, while holding other modes neutral ($|z|<1\sigma$). Only values with 95\% confidence intervals that exclude zero are shown. Confidence intervals were estimated from 1000-temporal block-bootstrap replicates using 12-month blocks.} 
\label{fig:4}
\end{figure}

The large ensemble of 100 KGAEs trained on Pacific basin SSTAs identified three modes of variability at distinct timescales. One mode varies on 10-year and longer timescales (decadal; Fig. \ref{fig:2}a,d), another on 3-7 year timescales (interannual; Fig. \ref{fig:2}b,e), and a third on 2-year timescales (quasibiennial; Fig. \ref{fig:2}c,f). Although the architecture includes five latent dimensions, two contribute little to the SSTA reconstruction variance due to the salient-variance weighting scheme in the loss function and are therefore excluded from the following analysis, which focuses on dominant modes of variability. The decadal, interannual, and quasibiennial modes show distinct spatial patterns (Fig. \ref{fig:3}a-c; Fig. \ref{fig:4}a-c) and time series (Figure \ref{fig:2}d-f). It should be noted that during hyperparameter tuning, architectures with different numbers of latent dimensions exhibited greater degrees of spectral overlap and, as a result, did not recover these same three modes as consistently as the final tuned architecture.

\begin{figure}[h]
\centering
\includegraphics[width=1\linewidth]{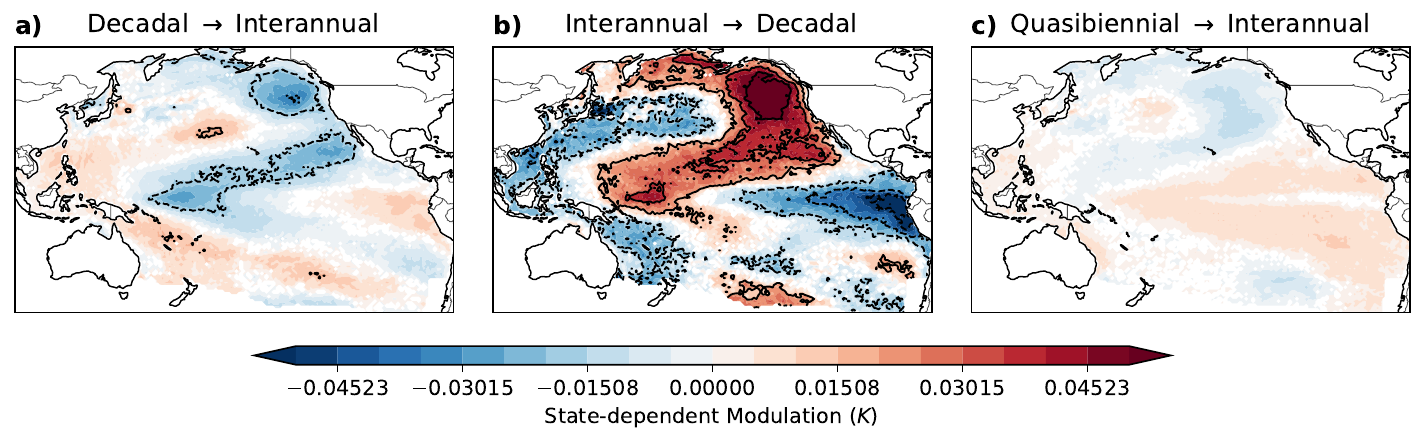}
\caption{a) Ensemble-mean ($n=100$) state-dependent modulation of the KGAE decoder responses, calculated using Equation \ref{statedependenteqn}. a) Modulation of the interannual mode response by the decadal mode. b) Modulation of the decadal-mode response by the interannual mode. c) Modulation of the interannual-mode response by the quasibiennial mode. Only values with 95\% confidence intervals that exclude zero are shown. Confidence intervals were estimated from 1000 i.i.d. bootstrap replicates generated by independently resampling the KGAE large-ensemble members.} 
\label{fig:5}
\end{figure}

\subsubsection{Decadal Mode}

The power spectral density of the KGAE decadal mode (Fig. \ref{fig:2}g) peaks at periods of 10-20 years and exhibits low-frequency variance characteristic of red noise. The temporal probability distribution of the decadal mode is strongly bimodal, suggesting two preferred phases (Fig. \ref{fig:2}a), although this bimodality is likely partially due to our choice of activation function (hyperbolic tangent). Two complementary approaches indicate that the linear spatial structure of the decadal mode yields consistent results. First, latent basis-vector finite-difference analysis shows that the associated spatial pattern of the decadal mode is consistent across the large ensemble members ($n=100$), and thus robust to random initialization and training stochasticity (Fig. \ref{fig:3}a). Second, 12-month-block bootstrapped composite analysis shows that most of the linear signal across the Pacific basin, except in the eastern tropical Pacific, is present in detrended ERA5 SSTAs during the interleaved, time-stratified cross-validation period and distinguishable from sampling noise at the 95\% confidence level (Fig. \ref{fig:4}a). 

To place the KGAE decadal pattern in the context of commonly studied modes of Pacific variability, we systematically compared the spatial patterns learned by the KGAE large ensemble with reference SSTA patterns associated with standard climate indices. Specifically, we first computed latitude-weighted spatial correlations between the reference patterns and the large ensemble-mean linear-response patterns (Fig. \ref{fig:3}a-c). We then computed spatial correlations between the reference patterns and the most robust state-dependent modulation patterns (Fig. \ref{fig:5}a-c). We further carried out these comparisons over three domains (Table \ref{tab:kgae_climate_panels}). This analysis suggests that the KGAE decadal mode most closely resembles the PDO spatial pattern both basin-wide ($r\approx 0.84$) and in the North Pacific ($r\approx0.92$), whereas in the tropics it more closely resembles the negative phase of the NPGO ($r\approx -0.88$).

These quantitative comparisons are consistent with the visual structure of the KGAE decadal mode. The North Pacific portion of the KGAE decadal mode closely resembles the canonical EOF structure typically associated with the PDO, and its time series captures recognizable PDO-related events, including the climate shifts of the late 1970s and the early 21st century (Fig. \ref{fig:2}d) \citep{miller_1976_climate_shift, mcphaden_21st_cent_climate_shift}. As in the canonical PDO pattern, cold SSTAs northwest of Hawaii extend westward into the Kuroshio Extension and are flanked by warm SSTAs along the west coast of North America. Consistent with this spatial similarity, the associated time series is positively correlated with the PDO index ($r\approx0.70$, other relevant correlations are listed in Table \ref{tab:Atable}, and KGAE latent time series overlaid with standard ENSO and PDO indices are shown in Supplemental Figure 3) \citep{mantua1999}. However, the traditional PDO index exhibits substantial interannual spectral power, whereas the KGAE decadal mode does not, suggesting that the KGAE decadal mode isolates a different subset of processes than those represented by standard PDO indices. 

The spatial footprint of the KGAE decadal mode also resembles other modes of variability identified in previous studies. Using a LIM approach, \citet{capotondi_optimal_2022} identified a mode termed the ``North Pacific-Central Pacific" (NP-CP) mode, with SSTAs spanning both the northeastern Pacific and the central equatorial Pacific \citep{capotondi_optimal_2022}. The KGAE decadal mode is also reminiscent of the Pacific Meridional Mode (PMM), defined through maximum covariance analysis by \citet{chiang_analogous_2004}, which exhibits a negative zonal SST gradient in the tropics and positive SSTAs extending northward toward the west coast of North America \citep{chiang_analogous_2004, richter_disentangling_2022}. In the South Pacific, the KGAE decadal mode captures cold anomalies stretching from the Maritime Continent to the southern coast of Chile, consistent with the South Pacific Meridional Mode \citep{you2018south}.

The decoder sensitivity analysis also shows a consistent and strong nonlinear response to the KGAE decadal mode, concentrated primarily in the North Pacific and western Pacific (Fig. \ref{fig:3}d). Temporal 12-month block-bootstrapped composites indicate that this signal is distinguishable from finite-sampling uncertainty in the ERA5 SSTAs, most clearly in the North Pacific and tropical western Pacific (Fig. \ref{fig:4}d). The decadal mode's self-dependence has a spatial pattern similar to, but opposite in sign from, its linear response pattern (Fig. \ref{fig:3}g). This relationship indicates that changes in the decadal latent coordinate produce progressively smaller changes in the decoded SSTA pattern as the magnitude of the latent coordinate increases. Together with the temporal bimodality of the decadal latent mode, this reduced decoder sensitivity suggests regime-like separation between its preferred phases, although the decoder analysis alone does not establish this interpretation. This reduced sensitivity at larger latent magnitudes could arise partly from our choice of activation function (hyperbolic tangent) or reflect learned differences between the mode's two preferred phases.

Although our composite methodology restricts the other modes to within one standard deviation of zero, it does not fully remove covariance among modes. Consequently, estimated nonlinear responses may reflect covariance with other modes in addition to nonlinearity in the decoder response to the decadal mode. More generally, the spatial correlations reported here are intended only as an approximate measure of pattern similarity. Since modes of variability are deeply entangled and often cross-correlated, the ``highest correlation" should be interpreted as a heuristic measure rather than a definitive mode identification.

\subsubsection{Interannual Mode}

The KGAE interannual mode exhibits a robust linear response, supported by both methods used to characterize its spatial pattern. The decoder sensitivity analysis consistently produces an ENSO-like spatial pattern, with pronounced warm SSTAs in the tropical Pacific and cold SSTAs in the North Pacific (Fig \ref{fig:3}b). Bootstrapped composites indicate that these signals are distinguishable from finite-sampling uncertainty in the detrended ERA5 SSTAs (Fig. \ref{fig:4}b). The time series of the interannual mode also captures several prominent ENSO events, including those of 1972-1973, 1982-1983, and 1997-1998 \citep{trenberth_definition_of_el_nino_1997}, and its power spectral density peaks at periods of 3-7 years \citep{hasselmann_stochastic_1976}. 

Pattern correlations further confirm that the KGAE interannual mode closely resembles standard ENSO patterns, including regression patterns based on Niño-3.4 ($r\approx0.81$) and the E- and C-indices ($r\approx0.80$ and $r\approx0.81$, respectively). The KGAE interannual mode is also temporally correlated with the Niño-3.4 index ($r\approx0.82$) \citep{barnston_documentation_1997}. At the same time, the somewhat weaker temporal correspondence and differences in the North Pacific pattern suggest that the interannual mode may isolate a timescale-constrained subset of variability only partly represented by traditional ENSO indices, or a different subset of North Pacific variability. In particular, cold SSTAs near the Aleutian Low are displaced eastward relative to standard ENSO regression patterns and appear more consistent with the atmospheric bridge mechanism, which links wintertime tropical warming to subsequent spring cooling in the North Pacific \citep{Alexander_AtmosphericBridge}. 

The decoder sensitivity analysis identifies a nonlinear response that is also supported by bootstrapped composites (Fig. \ref{fig:3}e and \ref{fig:4}e). This response broadly mirrors the linear-response pattern, with positive SSTAs extending along the South American coast and eastern equatorial Pacific, and a stronger cold anomaly in the Aleutian Low region. This asymmetry suggests amplification of the interannual mode pattern between positive and negative phases. However, the self-state-dependent nonlinearity of the interannual mode is of substantial magnitude and opposite in sign (Fig. \ref{fig:3}h). This pattern indicates that larger-amplitude latent states are associated with smaller marginal increases in the decoded spatial response. Together with the temporal bimodality of the latent mode, this reduced decoder sensitivity suggests regime-like separation between its preferred phases, although the decoder analysis alone does not establish this interpretation. Like for the decadal mode, this reduced sensitivity could be partly due to our choice of activation function (hyperbolic tangent), learned differences between the mode's preferred phases, or both. 

\subsubsection{Quasibiennial Mode}

The KGAE quasibiennial mode exhibits a robust and consistent linear response with a spatial pattern reminiscent of ENSO (Fig. \ref{fig:3}c), characterized by strong SSTAs along the equator and off the coast of South America, similar to a canonical El Niño \citep{Capotondi_Understanding_ENSO_diversity}. This spatial pattern is distinguishable from finite-sampling uncertainty in the ERA5 SSTAs (Fig. \ref{fig:4}c). Spatial correlations further support an association between the KGAE quasibiennial mode and eastern-Pacific ENSO variability: basin-wide, the quasibiennial mode is most strongly correlated with the E-Index regression pattern ($r\approx0.84$). Its power spectrum is concentrated at shorter timescales than those of standard ENSO indices (Fig. \ref{fig:2}g). The quasibiennial mode is moderately temporally correlated with the Niño-3.4 index ($r\approx0.65$) and, to a lesser extent, with the KGAE interannual mode ($r\approx0.49$), whereas the KGAE interannual mode is more strongly correlated with Niño-3.4 ($r \approx0.82$). That each KGAE mode is more strongly correlated with Niño-3.4 than with the other suggests that they capture distinct components of ENSO-related variability rather than two fully disentangled modes.

Both KGAE ENSO-like modes exhibit large magnitude SSTAs along the central and eastern equatorial Pacific (Fig. \ref{fig:3}b,c), but their spatial patterns differ in important ways. The quasibiennial mode displays a narrower equatorial band of SSTAs and remains more tightly confined to the eastern and central tropical Pacific. Differences in the extratropical response are particularly pronounced, with the two modes exhibiting anomalies of opposite sign in the North Pacific (Fig. \ref{fig:3}b,c; Fig. \ref{fig:4}b,c). For example, the quasibiennial mode shows a North Pacific dipole that is displaced relative to that of the interannual mode. The decoder sensitivity analysis (Fig. \ref{fig:3}f,i) and composite analysis (Fig. \ref{fig:4}f) both indicate that, unlike the interannual mode, the quasibiennial mode does not exhibit a robust nonlinear response.

\subsubsection{Seasonal Modulation of the Interannual and Quasibiennial Modes}

\begin{figure}[h]
\centering
\includegraphics[width=0.7\linewidth]{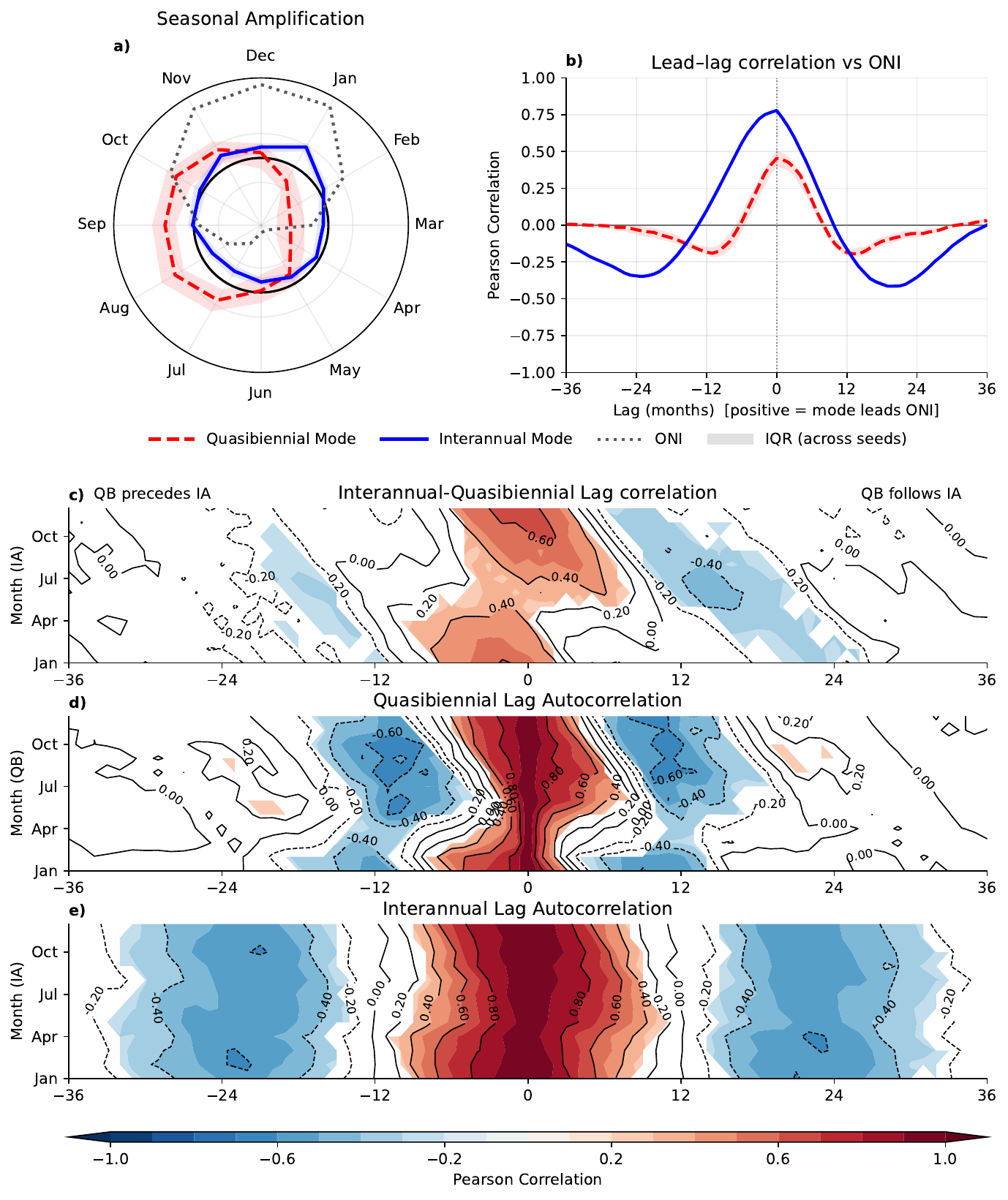}
\caption{a) Seasonal amplification of the KGAE quasibiennial and interannual modes and ONI, as legend indicated. Amplification is expressed as the percent departure of each climatological monthly-mean magnitude, $M_m$, from the annual-mean magnitude, $\overline{M}$: $100\times(|M_m|-\bar{|M|})/\bar{|M|}$. The black circle denotes zero amplification, with the inner and outer gray circles demarcating $-10\%$ and $+10\%$ increments, respectively. Shading indicates the interquartile range across the KGAE large ensemble. b) Lag correlations between the ONI and the KGAE quasibiennial and interannual modes. Positive lags indicate the KGAE mode leads ONI. c) Lag cross-correlations between the KGAE interannual and quasibiennial modes by calendar month (y-axis), with lag in months shown on the x-axis. Negative lags indicate that the quasibiennial mode leads. d) As in c), but for lag autocorrelations of the quasibiennial mode. e) As in d), but for lag autocorrelations of the interannual mode. Only statistically significant coefficients are shown ($p<0.05$, two-sided); no bootstrapping is used.}
\label{fig:6}
\end{figure}

The KGAE interannual and quasibiennial modes both exhibit pronounced seasonal cycles in amplitude, with a clear phase offset between their peak seasons (Fig. \ref{fig:6}a). We quantify this seasonal amplification as the percentage difference between the climatological monthly-mean magnitude and the corresponding annual-mean magnitude, relative to the annual mean. The quasibiennial mode is seasonally amplified during late boreal summer and fall (July-November), when ENSO events typically develop, whereas the interannual mode is amplified later, during late fall and boreal winter (November-January), when ENSO events typically mature. This phase relationship suggests that the quasibiennial mode is most active during ENSO development, while the interannual mode is most active during the mature phase of ENSO.

Lagged anomaly cross-correlations between the two modes provide additional support for this interpretation (Fig. \ref{fig:6}c). Peak positive correlations occur during November and December, when the two modes are most nearly in phase. During July-October, positive correlations at positive lags indicate that summer anomalies of the interannual mode are associated with same-sign anomalies of the quasibiennial mode during the following winter. During boreal spring, positive correlations at negative lags indicate that spring anomalies of the interannual mode are associated with the same-sign anomalies in the preceding winter quasibiennial mode. Together, these results suggest that anomalous activity in one mode tends to coincide with, lead, or follow anomalous activity of the same sign in the other mode. Negative correlations at longer positive lags (12-14 months) further suggest a tendency toward opposite-sign quasibiennial anomalies in the following year.

Taken together, these results suggest that the two modes capture distinct characteristics of ENSO variability. The quasibiennial mode appears to characterize the seasonal-to-biennial evolution of ENSO events, whereas the interannual mode is more closely associated with the mature-event amplitude of ENSO. This interpretation is also broadly consistent with monthly lag correlations between each mode and the Niño-3.4 index, both of which peak near zero lag (Fig. \ref{fig:6}b).

This interpretation is consistent with earlier work describing seasonal-to-biennial modulation of ENSO variability. One possible connection is the Tropospheric Biennial Oscillation (TBO), which involves coupled atmosphere-ocean-land interactions encompassing the Asian-Australian monsoon system and the tropical Pacific \citep{meehl1994coupled,loschnigg2003asian,li2001coupled,li2006spatiotemporal}. The interannual mode's relationship to ENSO amplitude and timing may also be consistent with the recharge-discharge mechanism of tropical Pacific ocean heat content \citep{jin_recharge_discharge_1997}. Meanwhile, the quasibiennial mode's timescale and spatial structure suggest possible ties to TBO-related Indo-Pacific coupling, as well as to equatorial Kelvin waves and off-equatorial Rossby wave dynamics \citep{meehl_coupled_2003, meehl_tropospheric_2001,meehl1994coupled,loschnigg2003asian,li2001coupled,li2006spatiotemporal,kirtman_rossby_1997}. We note that other analyses have identified ENSO growth and decay modes associated with seasonal-to-biennial timescale variability using Floquet analysis \citep{vimont_role_2022, thompson_linear_2000}, extended EOF analysis \citep{guan_pacific_2008}, and LIM \citep{shin_impact_2021}. We emphasize, however, that KGAE modes are descriptive in nature rather than inherently dynamical. The interpretation above concerns modulations of spatial patterns across timescales, not direct identification of physical mechanisms. The KGAE interannual and quasibiennial modes are therefore best described as two aspects of ENSO variability rather than as dynamical precursors or growth modes. 

Finally, lagged anomaly autocorrelations by month (Fig. \ref{fig:6}d,e) show that the interannual mode exhibits little seasonal variation in persistence, as expected due to the power spectra separation emphasized by the loss function, and an approximate decorrelation timescale of 24 months. In contrast, the quasibiennial mode shows a clear seasonal variation in persistence, consistent with ENSO's ``spring predictability barrier'' \citep{lopez_wwbs_2014, duan_spring_2013, mcphaden_tropical_2003}, and an approximate decorrelation timescale of one year. We note that these lagged autocorrelations are not simply a consequence of the seasonal cycle in mode amplitude.

\subsubsection{Nonlinear Interdependence of Interannual and Decadal Modes}

The relationship between interannual and decadal Pacific variability remains debated. Some studies have argued that part of decadal variability may arise as an artifact or integration of interannual variability \cite[e.g.,][]{Vimont_interannual_impact_on_decadal}, while others have identified decadal and longer-term modulations of ENSO and TBO dynamics \citep{Dieppois_ENSO_decadal_modulation, Sun_TPDV_on_ENSO_phase_frequencies, wang_interdecadal_nino_onset_1995,meehl2011decadal, an_inter-decadal_2016}. At the same time, ENSO itself exhibits substantial diversity in onset, zonal extent, and teleconnections, and is broadly partitioned into eastern-Pacific (EP) and central-Pacific (CP) types, although such classification depends on methodology \citep{Larkin_EP-CP_different_teleconnections, abdelkader_di_carlo_ENSO_diversity_definitions_controversy, Ashok_modoki, Freund_ENSO_diversity, karnauskas_canonical_vs_modoki, Kug_warmpool_coldpool_zonal_advective_feedback}. 

The decoder sensitivity analysis shows robust bi-directional nonlinear interactions between the KGAE interannual and decadal modes. The decoder response to the interannual mode is strongly modulated by the state of the decadal mode, and the decoder response to the decadal mode is similarly modulated by the state of the interannual mode  (Fig. \ref{fig:5}a,b). For the latter relationship (i.e., the change in the decoder response to the decadal mode between positive and negative states of the interannual mode), pattern correlations show that the modulation pattern is most strongly correlated with the PMM pattern across the full basin ($r\approx0.77$), in the tropics ($r\approx0.84$), and in the North Pacific ($r\approx0.7$). This is consistent with interpretations of the PDO as partly resulting from the integration of ENSO forcing through the atmospheric-bridge and re-emergence mechanisms \citep{Alexander_AtmosphericBridge, Deser_ENSO_PDV_CESM, Newman_PDO_revisited}, and with recent studies emphasizing coupling between the PMM and CP-type ENSO that contributes to decadal variability  \citep{stuecker_revisiting_2018, di2015enso, di_lorenzo_modes_2023}. Conversely, the modulation of the decoder response to the interannual mode by the decadal mode (Fig. \ref{fig:5}a) is also most closely related to the PMM, but with the opposite sign: it is robustly correlated with PMM regression map basin-wide ($r\approx-0.86$) and in the tropics ($r\approx-0.91$), while in the North Pacific it is more similar to the NPGO ($r\approx0.84$). Together, these results suggest that PMM-like structure provides the dominant statistical link between the interannual and decadal KGAE modes. 

\begin{figure}[h]
\centering
\includegraphics[width=1\linewidth]{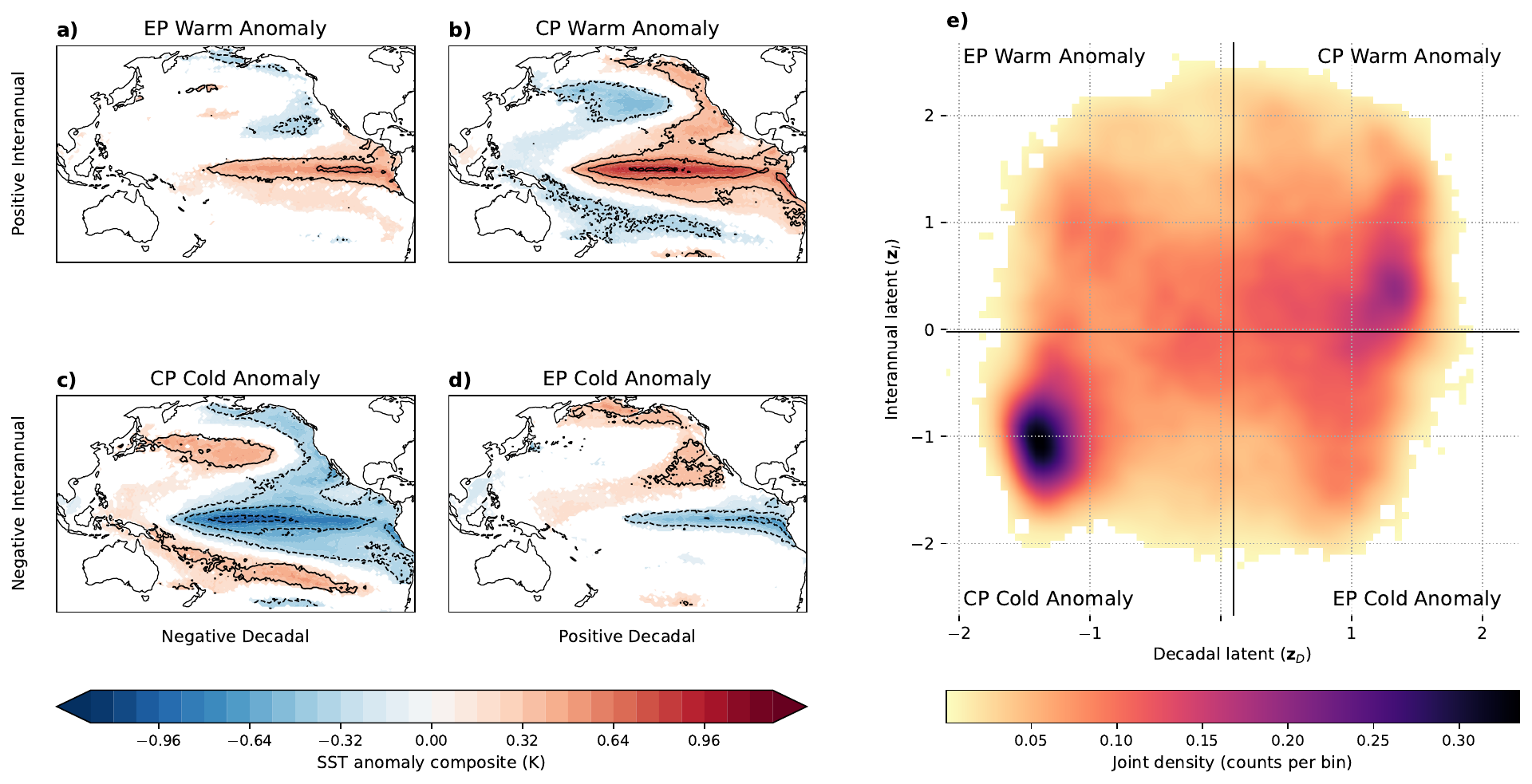}
\caption{a) Composite-mean monthly SSTAs for months when the ensemble-mean KGAE decadal mode is below its median ($z_{\mathrm{D}}<$ its median) and the ensemble-mean interannual mode is above its median ($z_{\mathrm{I}}>$ its median). b-d) The same as (a), but for b) the positive phase of both modes, c) the negative phase of both modes, and d) the positive decadal and negative interannual phases. Only values with two-sided 95\% confidence intervals that exclude zero are shown. Confidence intervals were estimated from 1,000 temporal block-bootstrap replicates using 12-month blocks. SSTA contours are at 0.3K intervals. e) Joint distribution of the KGAE decadal and interannual modes, pooled across the large-ensemble members and smoothed with a two-dimensional Gaussian kernel density estimator ($\sigma$=1.2 histogram bins). Black lines in (e) indicate the medians of ensemble-mean latent modes and divide the joint distribution into the four phase combinations used to construct the composites in panels (a-d). The upper-left, upper-right, lower-left, and lower-right quadrants correspond to panels (a), (b), (c), and (d), respectively.}
\label{fig:7}
\end{figure}

Figure \ref{fig:7} provides additional support for the interpretation that ENSO flavor may depend on the background state set by the KGAE decadal mode. When the decadal and interannual modes are both positive (Fig. \ref{fig:7}b), their effects reinforce one another and produce equatorial Pacific SSTAs resembling a mature CP-type ENSO event; the resulting pattern is strongly correlated with the C-Index pattern ($r\approx0.95$). When the interannual mode is positive and the decadal mode is negative, the two interfere, yielding weaker warm anomalies confined to the eastern tropical Pacific. Analogous zonal shifts occur for La Niña-like events: when both modes are negative (Fig. \ref{fig:7}c), the resulting pattern is strongly anticorrelated with the C-Index regression map ($r\approx-0.95$), whereas opposite-sign combinations produce weaker cold SSTAs confined to the eastern tropical Pacific (Fig. \ref{fig:7}d). Thus, same-sign combinations of the interannual and decadal modes most closely resemble CP-type ENSO, whereas opposite-sign combinations are only weakly reminiscent of EP-type ENSO. In this sense, interference between the interannual and decadal modes appears to influence ENSO flavor.

Several physical mechanisms could plausibly underpin such decadal and extratropical modulation of ENSO. North Pacific SSTAs can alter the Aleutian Low and Pacific High, thereby modulating tropical trade winds and the background state for coupled ENSO dynamics, including the Bjerknes feedback \citep{chen2023enhanced}. The PMM is associated with subtropical meridional SST gradients that induce pressure gradients and surface wind anomalies, amplifying the original SST gradients through wind-evaporation-SST feedback and favoring CP-type ENSO events \citep{yu_relationships_2011, yu_subtropics-related_2010, larson_pacific_2014, vimont_seasonal_2003, chang_pacific_2007}. Additionally, decadal modulation of the equatorial thermocline, such as the decreased thermocline slope associated with the positive PDO phase, can influence the zonal propagation of equatorial SSTAs and favor EP-type ENSO events \citep{vimont_optimal_2014, capotondi_optimal_2015}. We emphasize, however, that the KGAE modes are descriptive rather than dynamical. The decoder sensitivity analysis is only an approximation of the true decoder geometry, the network architecture is limited in capacity, and the training record is short. Accordingly, we do not interpret these results as establishing physical causality. Rather, the KGAEs suggest that the interannual and decadal modes together describe a two-dimensional continuum of ENSO-related variability, with each axis, by construction, constrained to a distinct timescale.

\subsection{Robustness of Learned Geometries}

\subsubsection{Sensitivity to Time-Stratified Cross-Validation Sampling Bias}

In the large-ensemble experiment, 100 randomly initialized KGAEs were trained for each of the five time-stratified cross-validation iterations. A unique seed was set for the pseudorandom number generation used to initialize model weights, determine minibatch ordering, and produce random samples during training. For each initialization, a corresponding ``full-period" KGAE was trained on all five folds using the same random initialization. Pairing the fold-specific and full-period models by initialization, and therefore random seed, allowed us to quantify sensitivity to the structured subsampling introduced by the time-stratified cross-validation scheme while reducing the influence of differences in random initialization and training stochasticity. For each fold, we computed the spatial correlation between the linear-response patterns from the fold-specific models and those from the corresponding full-period models. We then generated 1000 bootstrap replicates by independently resampling the 100 paired KGAE large ensemble members to estimate the sampling distribution of the ensemble-mean pattern correlations for each fold. These distributions are shown in Supplemental Figure 4. Overall, the learned large-scale patterns are highly robust to this structured subsampling. Pattern correlations are generally greater than $r\approx0.97$ for the quasibiennial mode, whereas the slower interannual and decadal modes exhibit even greater stability, with pattern correlations exceeding $r>0.995$.   

\subsubsection{Sensitivity to Ensemble Size}

We assess whether the ``large-ensemble" was sufficiently large to characterize the KGAE's learned latent-to-output geometry, as represented by its ensemble-mean decoder responses. For progressively larger candidate ensemble sizes $m$, we repeatedly drew pairs of non-overlapping subsets without replacement from our 100 randomly initialized KGAE ensemble members. We repeated this procedure 1,000 times for each value of $m$. For each pair of subsets, we calculated the spatial correlations between their ensemble-mean linear and state-dependent response patterns. Repeating this procedure yielded distributions of pattern correlations for each candidate value of $m$. We then examined how the lower tail ($5^{th}$ percentile) of the resulting pattern correlations varied with ensemble size $m$ (Supplemental Figure 5). 

The linear responses stabilized rapidly, with little additional change beyond $m=10$. The decadal and interannual linear responses were already stable at the smallest tested ensemble size $m=5$, while their prominent state-dependent responses stabilized around $m=20$. Most other nonlinear state-dependent responses stabilized before $m=50$, the largest subset size for which two non-overlapping subsets could be drawn from the 100 trained members, although some did not stabilize over the range examined. Overall, these results suggest that 100 members are sufficient to characterize the learned spatial patterns for this dataset, time period, and KGAE architecture.

We next used the large-ensemble experiment to evaluate KGAE out-of-sample generalization under systematic changes in the training period. Computational constraints precluded repeating the full 100-member experiment for each train/test split, so we sought to identify the minimum ensemble size that could approximate the large ensemble linear responses with acceptable stability. To do so, we leveraged the large ensemble experiment by repeatedly drawing subsets of size $m$ without replacement and computing the corresponding ensemble-mean linear responses. We then increased $m$ until 95\% of the subset-mean patterns exhibited spatial correlations greater than $r\approx 0.95$ with the full 100-member ensemble mean (Supplementary Figure 5). Under this criterion, an ensemble size of $m=5$ was sufficient to stably approximate the large-ensemble linear responses. 

\subsection{Assessing Distributional Shift with Progressively Increasing Splits}\label{exp2resultshere}

In the large-ensemble experiment, the held-out period of 2015-2023 limits the effective number of independent samples for the decadal and interannual modes. For example, a mode with a characteristic period of about five years may complete only one or two cycles over 2015-2023. To more robustly assess out-of-period generalization of these two modes, we repeated the analysis using nine progressively expanding contiguous train/test splits, each with an ensemble of five randomly initialized KGAEs ($n=5$). We began with a 50\%/50\% train/test split, corresponding to 1940/01-1982/01 for training (and validation) and 1982/02-2023/12 for testing. We then increased the training fraction in increments of 5\%, ending with a 90\%/10\% split corresponding to 1940/01-2015/08 for training (and validation) and 2015/09-2023/12 for testing.

For each split, we applied interleaved, time-stratified five-fold cross-validation within the combined training and validation period. During each time-stratified cross-validation iteration, four folds were used for training and one was held out for validation. Each KGAE was evaluated on its corresponding training and validation data, producing in-sample training-period and out-of-sample validation-period latent distributions. The held-out validation outputs were also reassembled across the five time-stratified cross-validation iterations to represent the complete validation period. We then trained a corresponding ensemble of five randomly initialized KGAEs ($n=5$) using nine progressively expanding continuous train/test splits. All five interleaved folds were used as training minibatches, with no fold held out for validation. This full-period KGAE ensemble generated out-of-sample latent modes for the corresponding contiguous test period.

\subsubsection{Linear Baseline Method}

We used principal components analysis (PCA) as a linear baseline for comparison. For basin-wide SSTAs, the first principal component (PC1) captures ENSO-like tropical Pacific variability, while the second (PC2) captures PDO-like North Pacific variability. We therefore treat PC1 and PC2 as linear baselines for the KGAE interannual and decadal modes, respectively. To enable direct comparison, we treated PCA (retaining only the first two components) as a ``linear autoencoder'' and fit it using the same preprocessing and interleaved, time-stratified cross-validation procedures as the KGAEs. EOF signs were aligned to the ONI regression map to ensure a consistent sign convention.

\subsubsection{Test-Set Distributional Shift}

We quantified test-set distributional shift for KGAE modes and principal components using the Wasserstein Distance (WD), which measures the amount of probability mass that must be transported to transform one distribution into another \citep{zbMATH03429908}. For each progressive split, we computed the WDs for two comparisons. The first compared the validation- and test-period distributions, providing a measure of out-of-period generalization. The second compared the in-sample training- and out-of-sample validation-period distributions, providing a measure of the distributional shift introduced by the time-stratified cross-validation scheme. Sampling uncertainty was estimated using a 1000 temporal block-bootstrap replicates with 12-month blocks. 

\begin{figure}[h]
\centering
\includegraphics[width=1\linewidth]{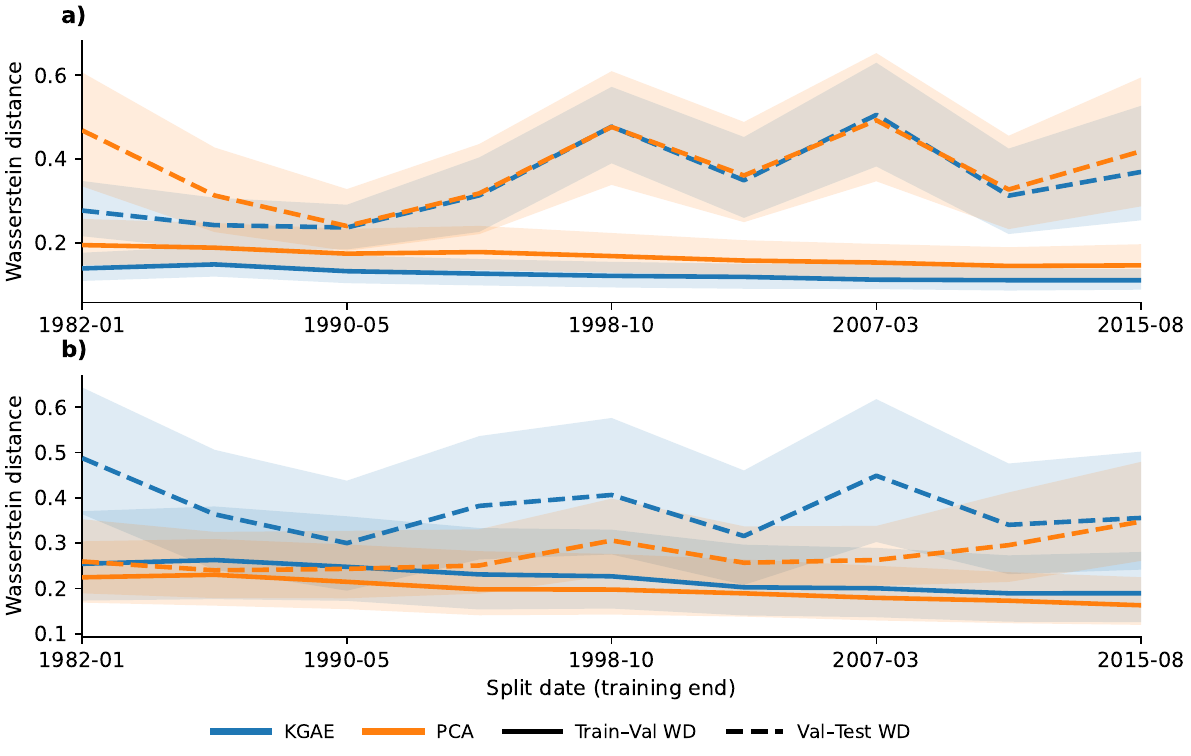}
\caption{a) Ensemble-mean WD between the validation- and test-period distributions (Val-Test; dashed lines) for the KGAE interannual mode (blue) and PC1 (orange) of basin-wide SSTAs across the nine contiguous train/test splits. Corresponding WDs between the training- and validation-set distributions are also shown (Train-Val; solid lines). Where training sets overlap between folds, the mean is taken across folds. Validation sets do not overlap between folds. b) Same as (a), but for the KGAE decadal mode and PC2. Shading denotes 95\% confidence intervals estimated from 1,000 temporal block-bootstrap replicates with 12-month blocks. The solid and dashed lines represent the medians of the same temporal block-bootstrap replicate distributions. }
\label{fig:8}
\end{figure}

Figure \ref{fig:8}a shows the bootstrapped WDs of the KGAE interannual mode and PC1 across the progressive splits. The two behave similarly, with overlapping confidence intervals at every split. Fig. \ref{fig:8}b shows the corresponding comparison for the KGAE decadal mode and PC2, with the same overall conclusion. Together, these results indicate that the KGAEs and PCA exhibit comparable out-of-period distributional shifts under systematic changes in the training period. 

The training- and validation-period WDs shown in Figure \ref{fig:8}a-b provide a complementary estimate of the distributional shift introduced by the time-stratified fold structure. This shift decreases as the combined training and validation period lengthens, supporting our use of a relatively long period (89\% of data) for the large-ensemble experiment. Across all splits, the KGAE and PCA confidence intervals overlap. 

\subsection{Evaluation of Variability in E3SM2 and CESM2}

Trained KGAEs and their latent representations can be used to evaluate variability in physics-based models. Here we assess quasibiennial-to-interannual and interannual-to-decadal variability in the Department of Energy's Energy Exascale Earth System Model version 2 \citep[E3SM2;][]{golaz2022doe} and the Community Earth System Model version 2 \cite[CESM2;][]{fasullo_E3SM_2024, rodgers2021ubiquity, danabasoglu_cesm2_2025}. We use KGAEs trained on ERA5 SSTAs to evaluate the historical runs of both models, restricted to 1940-2014 due to the lack of ERA5 reference data before that period. E3SM and CESM data are regridded to 1-degree equiangular grids using bilinear interpolation to match the ERA5 training data. Trends and climatologies were calculated separately for E3SM and CESM over the 1940-2014 historical period, and the resulting time series were removed following the same procedures described above to isolate biases in model internal variability from model drift and seasonality. SSTAs are then passed through the encoder to produce a set of KGAE latent posterior means for each model as defined by spatial patterns learned from ERA5. This procedure is done for each KGAE ensemble member.

\begin{figure}[h]
\centering
\includegraphics[width=0.9\linewidth]{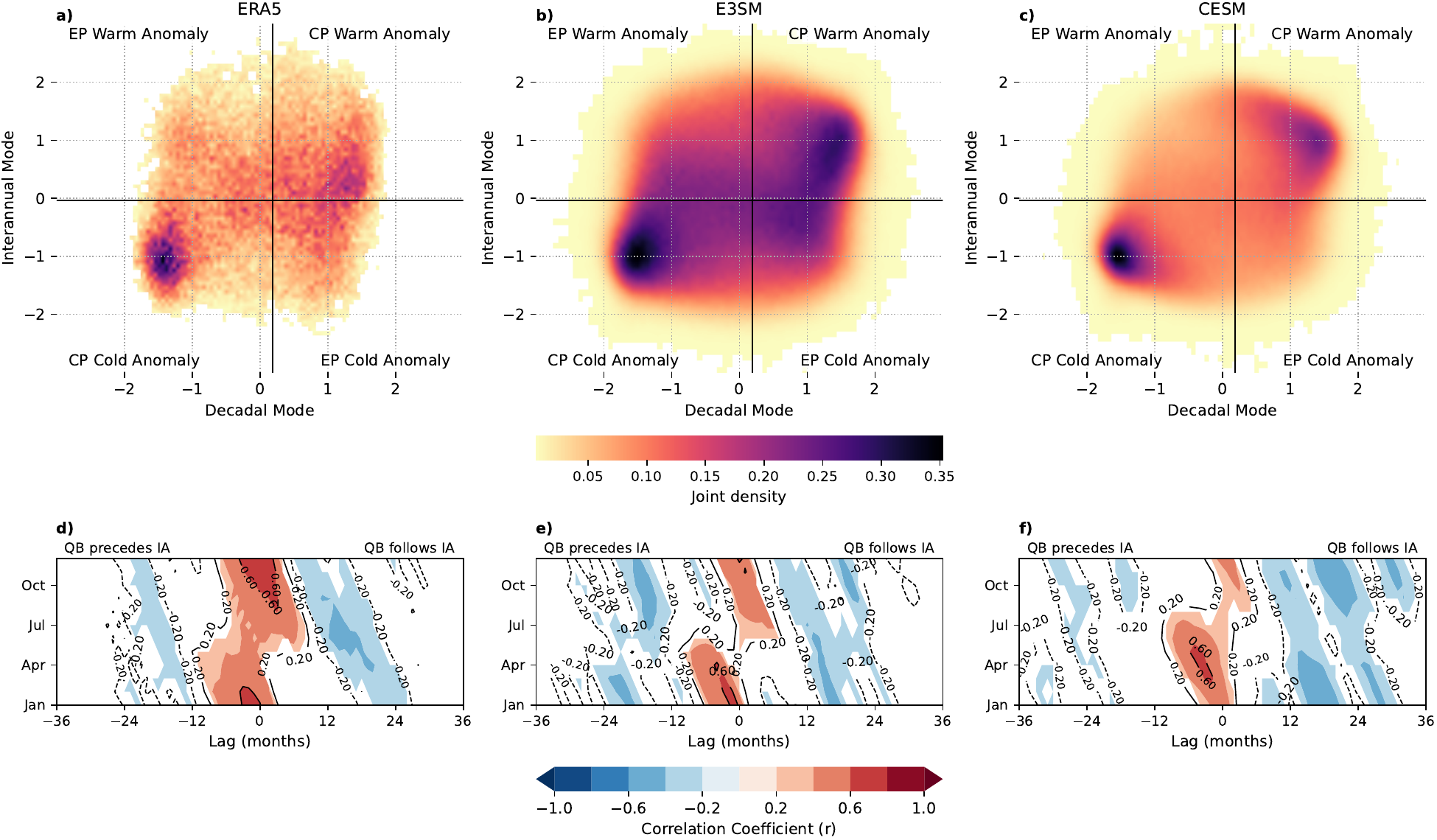}
\caption{a) Joint frequency histograms of KGAE decadal and interannual modes (1940-2014) from ERA5. The x- and y-axes indicate the decadal and interannual mode values at a given time. b-c) Same as a), but for E3SM2 and CESM2, respectively. d) Lag cross-correlations between the KGAE interannual and quasibiennial modes calculated from ERA5. Only correlation coefficients statistically significant at the $p<0.05$ level according to a two-sided significance test are shown; no bootstrapping is used. e-f) Same as d), but for E3SM2 and CESM2, respectively.}
\label{fig:9}
\end{figure}

Figure \ref{fig:9}a corresponds to Figure \ref{fig:7}e, the joint distribution of the ERA5-based KGAE decadal and interannual modes, reproduced here without Gaussian kernel smoothing for comparison. The four quadrants of the joint distribution correspond to different flavors of ENSO, with the upper-right and lower-left quadrants showing strong correspondence to mature CP-type El Niño and La Niña, respectively (Fig. \ref{fig:7} a-d). The upper-left and lower-right quadrants display anomaly patterns reminiscent of EP-type El Niño and La Niña events, respectively, although the correspondence is weaker than that of the other quadrants with CP-type events. ERA5 exhibits greater frequency in the lower-left quadrant, indicating a relative dominance of CP-type La Niña over EP-type La Niña (Fig. \ref{fig:9}a). The joint distribution is also more evenly spread between the top-right and bottom-right quadrants (positive KGAE decadal mode) than between the top-left and bottom-left quadrants, which is consistent with an interpretation of El Niño and La Niña as asymmetrical, with El Niño events displaying greater heterogeneity \citep{capotondi_enso_2020, Newman_ENSO_flavors, Capotondi_Understanding_ENSO_diversity}. The encoded E3SM2 SSTAs show frequencies comparable to those of ERA5, with a higher frequency of CP-type El Niño events and a lower frequency of EP-type La Niña events (Fig \ref{fig:9}b). Fig. \ref{fig:9}c reflects the same for CESM2, and displays a distinct bias towards more frequent CP-type El Niño and La Niña, consistent with the model's tendency to enhance ENSO variability in the central equatorial Pacific \citep{capotondi_enso_cesm}. We note that the smoother frequency distributions of E3SM2 and CESM2 may be partly due to the larger sample size of 21 ensemble members each, which have been pooled to construct a well-sampled joint distribution. Neither model fully captures the diversity of the ENSO continuum observed in ERA5, though E3SM2 is more aligned with ERA5 than CESM2, consistent with \citet{fasullo2024modes}.

Figures \ref{fig:9}d)-f) show interannual-quasibiennial monthly lag cross-correlations for ENSO in ERA5, E3SM2, and CESM2, respectively. Figure \ref{fig:9}d is equivalent to Figure \ref{fig:6}c, reproduced here for easier comparison. The lag cross-correlations show that neither E3SM2 nor CESM2 accurately captures the seasonal evolution of ENSO (Fig. \ref{fig:9}d-f). E3SM2 captures the timing of ENSO onset, development, and decay (albeit with weaker lag correlations), but does not capture the timing of the following event of opposite phase (Fig. \ref{fig:9}e). E3SM2 develops the following ENSO event in late boreal spring (April), whereas it develops in summer and autumn in ERA5. CESM2 exhibits more pronounced timing biases, including an ENSO phase transition in July and August, rather than April and May, as in ERA5 (Fig. \ref{fig:9}f). CESM2 also exhibits an anticorrelation starting at lag $+12$ to $+36$ months, especially after spring.

\section{Discussion}\label{sec5}

This study develops KGAEs (Knowledge-Guided AutoEncoders) and demonstrates their application in uncovering physically interpretable modes of Pacific variability across various timescales. Without imposing predefined filters or thresholds, KGAEs trained on Pacific basin SSTAs identified dominant decadal, interannual, and quasibiennial modes of variability. These learned modes exhibit structure consistent with known patterns, such as the PDO, flavors of ENSO, the TBO, and the PMM. The decoder sensitivity analysis further showed that KGAEs learn nonlinear state dependencies that characterize relationships across timescales. Complementary analyses using 12-month block-bootstrap replicates demonstrated that the inferred linear and nonlinear responses are distinguishable from finite-sampling uncertainty in ERA5. Together, the large-ensemble and progressive-split experiments enabled us to examine sensitivity to random initialization, training stochasticity, and the sub-sampling introduced by our time-stratified cross-validation procedure. The KGAEs exhibited distributional shifts comparable to those of principal component analysis both within the training and validation period and in out-of-period test data. 

The quasibiennial mode appears to interact with the interannual mode through a seasonal phase relationship, consistent with ENSO growth and decay. The interannual mode is modulated by the background state of the decadal mode, with the dependence of the interannual-mode spatial pattern on the decadal mode ($\beta_{i,d}$) strongly correlated with the C-Index regression pattern ($r\approx0.89$ ), suggesting a role in shaping central Pacific (CP) versus eastern Pacific (EP) ENSO flavor. Conversely, the interannual mode also modulates decadal variability in a way that resembles the PDO in the North Pacific ($r\approx 0.84$) and the PMM basin-wide and in the tropics ($r\approx0.79$ and $r\approx 0.84$ respectively). These results are largely consistent with existing dynamical interpretations developed with other methods \citep{chang_pacific_2007, Newman_ENSO_forced_PDO, Deser_ENSO_PDV_CESM, deser_tropical_extratropical_sst_1995, capotondi_optimal_2015}. 

We further demonstrated that trained KGAEs can be used to evaluate the distributional realism of physics-based climate models without retraining new KGAE ensembles on model data. This property is useful because retraining directly on model data would cause the KGAE to learn the physics-based model's own decadal, interannual, and quasibiennial variability, potentially obscuring biases relative to reanalysis and complicating mode separation based on KGAE loss-function criteria (e.g., minimizing spectral overlap). Our evaluation revealed biases in the representation of decadal, interannual, and quasibiennial modes of variability in both E3SM2 and CESM2. Although neither model is perfect, E3SM2 captures decadal variability and key characteristics of ENSO more comprehensively than CESM2. These results are consistent with \citet{fasullo2024modes}, which showed that CESM2 overestimates ENSO amplitude. Our results additionally indicate that CESM2 produces more frequent CP-type El Niño events than ERA5, with their occurrence varying with the KGAE decadal mode.

More broadly, the ability of KGAEs to separate modes associated with ENSO phase from those of ENSO growth and decay suggests potential value for long-lead ENSO prediction \cite[12+ months;][]{petrova_long_lead_enso, ham_longlead_enso}. A preliminary analysis revealed that the spring predictability barrier is concentrated in the quasibiennial mode and largely absent from the interannual mode (Figure \ref{fig:6}d,e). This finding is notable because the predictability barrier is often associated with TBO-like transitions in the coupled Asia-Pacific system \citep{meehl_coupled_2003, meehl_tropospheric_2001,meehl1994coupled}. While prediction applications remain beyond the scope of this study, these results suggest that time-stratified latent decompositions may offer a useful framework for separating more predictable and less predictable components of ENSO-related variability.

\subsection{Limitations}

The KGAE method is subject to limitations. First, the knowledge-guided loss function explicitly prioritizes spatiotemporal disentanglement over maximizing the variance explained by the latent representation. KGAEs herein capture approximately 20\% of SSTA variance, whereas the first two principal components capture approximately 30\% (Supplemental Figure 7). This reduction in explained variance reflects the KGAE objective, which prioritizes separating timescale-dependent structure over maximizing variance explained. Second, the loss function relies on standard FFTs to estimate latent power spectra, which introduces several practical and conceptual constraints. Since FFT-based spectra require evenly sampled sequential data, each training minibatch must span sufficiently long analysis periods to resolve low-frequency variability. In our case, this requirement motivated the interleaved, time-stratified cross-validation scheme, which in turn introduces structured sub-sampling. Although a non-uniform FFT might relax the requirement of regular spacing, each minibatch would still need to sample the full record to retain sensitivity to slow variability. 

More fundamentally, interpreting a single FFT-based power spectrum as representative of the full record carries assumptions of stationarity that are not strictly satisfied by Earth system variability. The finite-record calculation also implicitly treats the series as periodic, making it sensitive to discontinuities at record boundaries. We attempted to mitigate nonstationarity issues by detrending SSTAs, but the second-order polynomial fit to the Pacific basin mean does not remove changes in SSTA covariance structure through time. Test-set spatial MSE fields also revealed coherent error structure, particularly in the Kuroshio-Oyashio Extension and along the equator (not shown). During model development, we applied split-cosine-bell tapers to the latent posterior means before computing power spectral density curves to reduce artifacts associated with periodicity, but these changes had little effect and were not retained in the final design. More generally, the present KGAEs do not explicitly represent time-varying spectral power, although some aspects of temporal modulations may be absorbed indirectly through nonlinear decoder responses. Explicit treatment of time-varying oscillatory power, as well as separation of internal variability from forced change, is left to future work. 

There are also interpretive limitations. Our interpretation of the learned latent modes is inferred through post hoc analyses based on statistical correlations and composites, which, while suggestive of physical links, do not establish causality. Likewise, the decoder sensitivity analysis is intended as a simple finite-difference diagnostic approximation of decoder behavior rather than a complete characterization of decoder geometry. Additionally, the bimodality and reduced decoder sensitivity at larger magnitudes of the decadal and interannual modes may be partially attributable to the hyperbolic tangent activation function used here. The quasibiennial mode displays a unimodal distribution, showing that the hyperbolic tangent activation function does not necessitate bimodality in the distributions of all modes, but it may contribute to or enhance bimodality where present. A simple box-average analysis of SSTA in the regions identified as centers of action identified for each mode (Supplemental Figure 8) shows that unfiltered regional SSTAs do not display the same strong bimodality as the KGAE latent modes. Thus, while KGAEs identify physically meaningful modes of variability, the characteristics of these modes may be influenced by architecture choices. Finally, the KGAEs presented herein exclusively study SSTA in the Pacific basin, which limits their ability to resolve modes of the full global climate system. The method could be expanded to be multivariate or global, but that is left for future work. Accordingly, we do not consider the view of Pacific SSTA variability provided by KGAEs as comprehensive, but rather as complementary to the perspectives offered by traditional methods.

\section{Code Availability}

The code used in this study is available at \url{https://github.com/kjhall01/kgae}. The xESMF Python library was used for data regridding \citep{zhuang_pangeo-dataxesmf_2025}. The KGAE method was implemented with PyTorch \citep{pytorch}. Hyperparameter tuning was conducted using the Optuna library \citep{akiba2019optuna}. 

\section{Author Contributions}

K.J.C.H. formulated the KGAE concept. K.J.C.H. and M.J.M. co-developed the KGAE and experimental methodologies. K.J.C.H. implemented the requisite software, and M.J.M. performed code review. K.J.C.H. and M.J.M. wrote the manuscript, and E.F.W., G.A.M., and A.C. provided critically important editorial contributions and feedback. 

\section{Competing Interests}

All authors declare no competing interests.

\clearpage
\acknowledgments
This material is based upon work supported by the U.S. DOE, Office of Science, Office of Biological and Environmental Research (BER), RGMA component of the Earth and Environmental System Modeling Program under Award \#DE-SC0024093. Portions of this work were also supported by the Regional and Global Model Analysis (RGMA) component of the Earth and Environmental System Modeling Program of the U.S. Department of Energy's Office of Biological and Environmental Research (BER) under award no. DE-SC0022070. Additionally, this work was supported by the National Center for Atmospheric Research, which is a major facility sponsored by the National Science Foundation (NSF) under Cooperative Agreement No. 1852977. This material is based upon work supported by the National Science Foundation Graduate Research Fellowship Program under Grant No. DGE 2236417. Any opinions, findings, and conclusions or recommendations expressed in this material are those of the author(s) and do not necessarily reflect the views of the National Science Foundation. A.C. was supported by the NOAA Climate Program Office Climate Variability and Predictability Program Award No. NA24OARX431C0024-T1-01. The use of Grammarly (\url{https://app.grammarly.com/}), GitHub Copilot (\url{https://github.com/features/copilot}), and ChatGPT (\url{https://chat.openai.com/}) is acknowledged to refine academic language and assist with code development. All AI-generated content has been reviewed and edited to ensure accuracy, and full responsibility is taken for the final content of this manuscript.

%
%
\datastatement
PDO index is available from NOAA National Centers for Environmental Information at \url{https://www.ncei.noaa.gov/access/monitoring/pdo/}. Oceanic Niño Index is available from NOAA Climate Prediction Center at \url{https://origin.cpc.ncep.noaa.gov/products/analysis\_monitoring/ensostuff/ONI\_v5.php}. ERA5 data are available through the Copernicus Climate Change Service \citep{Copernicus_Climate_Change_Service2019-op}. E3SM2 and CESM2 data are available through the National Energy Research Scientific Computing Center Portal (\url{https://portal.nersc.gov/archive/home/c/ccsm/www/E3SMv2/FV1/atm/proc/tseries/month\_1}) and Earth System Grid (e.g., \url{https://aims2.llnl.gov/search/?project=E3SM/}, and \url{https://aims2.llnl.gov/search/?project=CESM/}). CESM2 data are also available through the US NSF NCAR Geoscience Data Exchange (\url{https://gdex.ucar.edu/datasets/d651056/}).

%

\appendix[A]
\appendixtitle{Knowledge-Guided Autoencoder Methodology}
\label{appendix_a}
An autoencoder consists of an encoder ($E$), a decoder ($D$), and a latent space ($L$) \citep{michelucci2022introduction}. $L$ contains learned properties of data $X$ that enable reconstruction by $D$. Let $X \in \mathbb{R}^{N \times M}$ denote the full dataset, where $N$ is the number of samples (e.g., monthly SSTAs), and $M$ is the number of grid cells in the Pacific Ocean. The encoded latent representation is

\begin{equation}
    E(X) = L,
\end{equation}

\noindent where $L \in \mathbb{R}^{N \times C}$ is the latent space of dimensionality $C$, with $C \ll M$. The decoder maps this latent space back to the input space:

\begin{equation}
    D(E(X)) = D(L) = \hat{X},
\end{equation}

\noindent where $\hat{X}$ is the reconstructed version of $X$. In our application, $E$ and $D$ are fully connected neural networks with mirrored architectures. The latent activations for a mini-batch $\tilde{X} \in \mathbb{R}^{B \times M}$ are denoted by $Z = E(\tilde{X}) \in \mathbb{R}^{B \times C}$. A summary of notation is provided in Table \ref{taba1}.

\begin{table}[h]
    \centering
    \begin{tabular}{c|l|l}
        \textbf{Symbol} & \textbf{Definition} & \textbf{Property} \\
        \hline
        $E$ & Encoder neural network & $E(X)=L$ \\
        $D$ & Decoder neural network & $D(Z)=\hat{X}$ \\
        $X$ & All input data & $X \in \mathbb{R}^{N \times M}$ \\
        $\tilde{X}$ & Mini-batch of input data & $\tilde{X} \in \mathbb{R}^{B \times M}$ \\
        $\hat{X}$ & Reconstructed mini-batch & $\hat{X} = D(E(\tilde{X})) \in \mathbb{R}^{B \times M}$ \\
        $N$ & Total number of samples & \\
        $B$ & Mini-batch size & $B \ll N$ \\
        $M$ & Input feature dimension & \\
        $C$ & Number of latent neurons (i.e., bottleneck) & $C \ll M$ \\
        $L$ & Encoded latent space & $L \in \mathbb{R}^{N \times C}$ \\
        $\mu$ & Latent Posterior Means for $\tilde{X}$ & \\
        $\sigma^2$ & Latent Posterior Variance for $\tilde{X}$ & $\mathcal{N}(\mu, \sigma^2I) = E(\tilde{X}) \in \mathbb{R}^{B \times 2C}$ \\
        $\epsilon$ & I.I.D. isotropic Gaussian noise & $\epsilon \sim \mathcal{N}(0, I)$ \\
        $Z$ & Latent activations for $\tilde{X}$ & $Z = \mu + \sigma\odot\epsilon \in \mathbb{R}^{B \times C}$ \\
        $z_j$ & Activations of neuron $j$ across batch & $z_j = Z_{:,j} \in \mathbb{R}^{B}$ \\
        $\mathcal{F}$ & Discrete Fourier transform (DFT) & $\mathcal{F}[z_j](f_k) \in \mathbb{C}$ \\
        $f_k$ & Discrete frequency bin & $k = 1, \dots, K$ \\
        $\hat{P}_j(f_k)$ & Normalized power spectrum of $z_j$ & $\hat{P}_j(f_k) = \frac{|\mathcal{F}[z_j](f_k)|^2}{\sum_{k'=1}^{K} |\mathcal{F}[z_j](f_{k'})|^2}$ \\
        $W$ & Decoder first-layer weights & $W \in \mathbb{R}^{C \times H}$ \\
        $H$ & Width of decoder first hidden layer & \\
        $m_{ij}$ & Relative saliency of neuron $j$ for sample $i$ & Fraction of variance gradient attributable to $j$ \\
        $\mathrm{Corr}_{jj'}$ & Spatial correlation between decoded patterns of latent neurons $j$ and $j'$ & $\mathrm{Corr}_{jj'} = \mathrm{corr}(D_j,D_{j'})$ \\
        $v_j$ & Salient variance associated with latent neuron $j$ & Saliency-weighted reconstruction variance \\
        $s_j$ & Normalized saliency weight for latent neuron $j$ & $s_j = \frac{v_j}{\sum_{j'=1}^{C} v_{j'}}$ \\
    \end{tabular}
    \caption{Notation used throughout the paper.}
    \label{taba1}
\end{table}

The Knowledge-Guided AutoEncoder (KGAE) is trained to minimize a composite loss function composed of six additive terms:

\begin{equation}
    \text{Loss}_{\text{KG}} = \text{Term}_1 + \text{Term}_2 + \text{Term}_3 + \text{Term}_4 + \text{Term}_5 + \text{Term}_6.
\end{equation}

\noindent Each term contributes to different properties of the learned representations and is defined in the subsections that follow.

\subsection{Term 1: Reconstruction Loss}

The first term in $\text{Loss}_{\text{KG}}$ is the reconstruction loss. We use the mean squared error (MSE) to quantify the difference between the mini-batch inputs $\tilde{X} \in \mathbb{R}^{B \times M}$ and their reconstructions $\hat{X}= D(E(\tilde{X})) \in \mathbb{R}^{B \times M}$. This term encourages the autoencoder to faithfully reproduce the input data.

\begin{equation}
\text{Term}_1 = \mathrm{MSE}(\tilde{X}, \hat{X}) = \frac{1}{B M} \sum_{i=1}^{B} \sum_{j=1}^{M} \left( \tilde{X}_{ij} - \hat{X}_{ij} \right)^2.
\end{equation}

\subsection{Term 2: Spectral Overlap and Spatial Correlation}

The second term penalizes redundant latent representations based on spectral overlap and spatial correlation. It is defined pairwise over all latent neurons $j, j' \in \{1, \dots, C\}$ with $j \ne j'$ and weighted by saliency-based variance. 

The normalized power spectrum of neuron $j$ is

\begin{equation}
    \hat{P}_j(f_k) = \frac{\left|\mathcal{F}[z_j](f_k)\right|^2}{\sum_{k'=1}^{K} \left|\mathcal{F}[z_j](f_{k'})\right|^2},
\end{equation}

\noindent where $z_j = Z_{:,j} \in \mathbb{R}^{B}$ is the activation vector of neuron $j$ across the mini-batch and $f_k$ is the $k$th frequency bin.

For each pair $(j, j')$, compute the element-wise intersection and union of their normalized spectra:

\begin{equation}
    I_{f_k}^{(j,j')} = \min\left(\hat{P}_j(f_k), \hat{P}_{j'}(f_k)\right), \quad
    U_{f_k}^{(j,j')} = \max\left(\hat{P}_j(f_k), \hat{P}_{j'}(f_k)\right),
\end{equation}

\begin{equation}
    \text{IoU}_{jj'} = \frac{\sum_{k=1}^{K} I_{f_k}^{(j,j')}}{\sum_{k=1}^{K} U_{f_k}^{(j,j')}}.
\end{equation}

To compute saliency weights, we define the contribution of latent neuron $j$ to the reconstruction of sample $i$ as:

\begin{equation}
    m_{ij} = \frac{\left\lVert \frac{\partial \hat{\mathbf{X}}_{i,:}}{\partial Z_{ij}} \right\rVert_2}{\sum_{j'=1}^{C} \left\lVert \frac{\partial \hat{\mathbf{X}}_{i,:}}{\partial Z_{ij'}} \right\rVert_2}.
\end{equation}

The salient variance for neuron $j$ is then:

\begin{equation}
    v_j = \frac{1}{B} \sum_{i=1}^{B} m_{ij} \sum_{m=1}^{M} \left( \hat{X}_{im} - \bar{\hat{X}}_m \right)^2,
\end{equation}

\noindent where $\bar{\hat{X}}_m= \frac{1}{B} \sum_{i=1}^{B} \hat{X}_{im}$. We normalize across latent neurons:

\begin{equation}
    s_j = \frac{v_j}{\sum_{j'=1}^{C} v_{j'}}.
\end{equation}

The weighting for each pair $(j, j')$ is defined as:

\begin{equation}
    w_{jj'} =
    \begin{cases}
    s_j s_{j'} & \text{(multiplicative scheme)} \\
    s_j + s_{j'} & \text{(additive scheme)}
    \end{cases}
\end{equation}

\noindent with $w_{jj} = 0$, and normalize all off-diagonal values:

\begin{equation}
    w_{jj'} \leftarrow \frac{w_{jj'} \cdot (1 - \delta_{jj'})}{\sum_{j \ne j'} w_{jj'}},
\end{equation}

\noindent where $\delta_{jj'}$ is the Kronecker delta. To assess spatial similarity, we define $D_j \in \mathbb{R}^{M}$ as the decoded pattern obtained by decoding a latent basis vector with unit activation in neuron $j$ and zero activation in all other latent neurons. Then,

\begin{equation}
    \text{Corr}_{jj'} = \mathrm{corr}(D_j, D_{j'}).
\end{equation}

The complete Term$_2$ is then:

\begin{equation}
    \text{Term}_2 = \sum_{j=1}^{C} \sum_{j'=1}^{C} \text{IoU}_{jj'} \cdot (1 + |\text{Corr}_{jj'}|) \cdot w_{jj'}.
\end{equation}

\subsection{Term 3: Spectral Modality}

The third term encourages unimodal, bell-shaped spectral densities for each latent neuron. For each $z_j$ compute the normalized power spectrum:

\begin{equation}
    \hat{P}_j(f_k) = \frac{|\mathcal{F}[z_j](f_k)|^2}{\sum_{k'=1}^{K} |\mathcal{F}[z_j](f_{k'})|^2}.
\end{equation}

The cumulative sum of the normalized power spectrum is then

\begin{equation}
    F_j(f_k) = \sum_{k'=1}^{k} \hat{P}_j(f_{k'}).
\end{equation}

\noindent where $F_j(f_k)$ is the cumulative power up to frequency $f_k$ for neuron $j$, $\hat{P}_j(f_{k'})$ is the normalized power at frequency $f_{k'}$, $k'$ is the summation index, and $k$ is the frequency index. Next, a fourth-degree polynomial $\Phi_j(f_k)$ is fit in frequency space to the logit-transformed cumulative power spectrum as follows:

\begin{equation}
    \log\left( \frac{F_j(f_k)}{1 - F_j(f_k)} \right) \approx \Phi_j(f_k) = \sum_{p=0}^{4} \beta_{jp} f_k^p,
\end{equation}

\noindent where the logit transformation maps $F_j(f_k)$ from the unit interval to the real line, facilitating polynomial regression. The coefficients of the polynomial are denoted by $\beta_{jp}$, where $p$ ranges from 0 to 4. A bell-shaped density is recovered by applying the derivative of the logistic (sigmoid) function to the fitted polynomial:

\begin{equation}
    \tilde{P}_j(f_k) = \sigma\left(\Phi_j(f_k)\right) \cdot \left(1 - \sigma\left(\Phi_j(f_k)\right)\right),
\end{equation}

\noindent where $\sigma(\cdot)$ is the sigmoid function. The reconstructed spectrum is then normalized:

\begin{equation}
    \tilde{P}_j(f_k) = \frac{\tilde{P}_j(f_k)}{\sum_{k'=1}^{K} \tilde{P}_j(f_{k'})},
\end{equation}

\noindent allowing a valid comparison with the original normalized power spectrum $\hat{P}_j(f_k)$. The mean absolute error (MAE) between the true and reconstructed spectra is then computed across all neurons:

\begin{equation}
    \text{Term}_3 = \frac{1}{C\,K} \sum_{j=1}^{C} \sum_{k=1}^{K} \left| \hat{P}_j(f_k) - \tilde{P}_j(f_k) \right|.
\end{equation}

\subsection{Term 4: Centering}

The fourth term penalizes non-zero mean activations in the latent space across each mini-batch. Let $\mu_{ij}$ denote the latent posterior mean of neuron $j$ for sample $i$. The centering loss is given by:

\begin{equation}
    \text{Term}_4 = \frac{1}{C} \sum_{j=1}^{C} \left( \frac{1}{B} \sum_{i=1}^{B} \mu_{ij} \right)^2.
\end{equation}

\subsection{Term 5: Regularizing}

The fifth term regularizes the Frobenius norm of the decoder's first layer weight matrix $W \in \mathbb{R}^{C \times H}$, where $H$ is the width of the first decoder hidden layer. This term encourages the overall magnitude of the first-layer decoder weights to remain close to one and is evaluated at each training step. It is defined as

\begin{equation}
    \text{Term}_5 = \lambda \left| 1 - \left\| W \right\|_F \right|
    = \lambda \left| 1 - \sqrt{ \sum_{j=1}^{C} \sum_{h=1}^{H} W_{jh}^2 } \right|.
\end{equation}

where $\lambda =1e^{-4}$ is used as a weighting coefficient to de-emphasize regularization. 

\subsection{Term 6: Kullback-Leibler Divergence}

The Kullback-Leibler Divergence loss term is defined as the divergence between the approximate posterior over latent variables and the isotropic Gaussian prior. For a single sample:

\[
D_{\mathrm{KL}}\!\left(q_{\phi}(\mathbf{z} \mid \mathbf{x})\,\|\,p(\mathbf{z})\right)
=
\int q_{\phi}(\mathbf{z} \mid \mathbf{x})\,
\log\!\left(\frac{q_{\phi}(\mathbf{z} \mid \mathbf{x})}{p(\mathbf{z})}\right)\,dz.
\]

In our framework, where

\[
q_{\phi}(\mathbf{z} \mid \mathbf{x}) = \mathcal{N}\!\bigl(\boldsymbol{\mu},\mathrm{diag}(\boldsymbol{\sigma}^2)\bigr),
\qquad
p(\mathbf{z})=\mathcal{N}(\mathbf{0},\mathbf{I}),
\]

\noindent this term is implemented as

\[
\text{Term}_6 
= 
\beta\,D_{\mathrm{KL}}\!\left(q_{\phi}(\mathbf{z} \mid \mathbf{x})\,\|\,p(\mathbf{z})\right)
=
\frac{\beta}{2}
\sum_{j=1}^{C}
\left(
\sigma_j^2
+
\mu_j^2
-
1
-
\log \sigma_j^2
\right),
\]

\noindent where we use $\beta=0.01$ and average over a minibatch.

\subsection{Bootstrapping Methodology}
\label{app:bootstrap}

We assess the robustness of the conditional composites used in our analysis to sampling uncertainty by constructing confidence intervals using 12-month temporal block-bootstrap resampling. To construct such an interval for a conditional mean
\[
\mu = \mathbb{E}[X \mid C],
\]
where $C$ denotes the subset of samples satisfying a given condition. We generate an empirical distribution of $N$ bootstrap estimates, $p(\hat{\mu}_b)$, by repeatedly resampling contiguous 12-month blocks from the observed samples with replacement, and then applying the condition $C$. The number of samples used to compute each bootstrap estimate $\hat{\mu}_b$ is held equal to the number of observed samples. A $(1-\alpha)$-level bootstrap confidence interval for $\mu$ is then obtained from the $\alpha/2$ and $1-\alpha/2$ quantiles of $p(\hat{\mu}_b)$. We consider $\mu$ to be statistically distinguishable from zero at significance level $\alpha$ when this confidence interval excludes zero.

\subsection{Conditional Composite Finite-Difference Methodology}
\label{app:composite}

We use a conditional composite-based finite-difference method to estimate the linear (symmetric) and nonlinear (asymmetric) responses of SSTAs to various indices. We interpret the linear response of SSTA to index $z_j$ as the mean impact on the conditional composite average SSTA of an increase in $z_j$, while holding all other indices under consideration $z_{j'\neq j}$ within $1\sigma_{j'}$ of their mean values (note that Fig. \ref{fig:7} instead uses the condition ``greater than median value," with other modes unrestrained). Limited sample size requires the use of coarse finite difference estimates, derived by splitting $X$ into positive, negative, and neutral regimes: 

\[
\bar{X}_+ = \mathbb{E}[X \mid (z_j > \sigma_j), \ (|z_{j' \neq j}| < \sigma_{j'})],
\]
\[
\bar{X}_- = \mathbb{E}[X \mid (z_j < -\sigma_j), \ (|z_{j' \neq j}| < \sigma_{j'})],
\]
\[
\bar{X}_0 = \mathbb{E}[X \mid (|z_j| < \sigma_j), \ (|z_{j' \neq j}| < \sigma_{j'})].
\]

We then estimate the symmetric linear response as
\[
\mathrm{L}
=
(\bar{X}_+ - \bar{X}_0) + (\bar{X}_0 - \bar{X}_-)
=
\bar{X}_+ - \bar{X}_-
\]

We interpret the nonlinear (asymmetric) response of SSTA to $z_j$ as the difference between the impact of moving from the neutral regime to the positive regime and the impact of moving from the negative regime to the neutral regime. We estimate it as
\[
\mathrm{NL}
=
(\bar{X}_+ - \bar{X}_0) - (\bar{X}_0 - \bar{X}_-)
=
\bar{X}_+ + \bar{X}_- - 2\bar{X}_0
\]

During bootstrapping, the time series used to estimate each of $\bar{X}_+$, $\bar{X}_0$, and $\bar{X}_-$ is resampled independently $N$ times by drawing contiguous 12-month blocks with replacement from the original, unfiltered chronological time series until each bootstrap sample matches the original sample length. Appropriate conditions are applied to each bootstrapped replicate to produce estimates of $\bar{X}_+$, $\bar{X}_0$, and $\bar{X}_-$. These estimated quantities are then used to construct confidence intervals for both the linear and nonlinear responses. 

\clearpage 

\begin{table}[h]
    \centering
    \begin{tabular}{c|ccccccccc}
    \toprule
     & DM & IA & QB & NPGO & Niño3.4 & E-Index & C-Index & PMM & PDO \\
    \midrule
    DM & 1.00 & - & - & - & - & - & - & - & - \\
    IA & 0.27 & 1.00 & - & - & - & - & - & - & - \\
    QB & 0.09 & 0.49 & 1.00 & - & - & - & - & - & - \\
    NPGO & -0.42 & -0.11 & -0.09 & 1.00 & - & - & - & - & - \\
    Niño3.4 & 0.41 & 0.82 & 0.65 & -0.26 & 1.00 & - & - & - & - \\
    E-Index & -0.02 & 0.45 & 0.44 & 0.01 & 0.36 & 1.00 & - & - & - \\
    C-Index & 0.55 & 0.69 & 0.48 & -0.35 & 0.86 & -0.00 & 1.00 & - & - \\
    PMM & 0.57 & -0.14 & -0.27 & -0.44 & -0.04 & -0.57 & 0.32 & 1.00 & - \\
    PDO & 0.70 & 0.36 & 0.21 & -0.15 & 0.38 & 0.20 & 0.36 & 0.26 & 1.00 \\
    \bottomrule
    \end{tabular}
    \caption{Time-series correlation matrix, where DM indicates the ensemble mean KGAE decadal mode, IA indicates the ensemble mean KGAE interannual mode, and QB indicates the ensemble mean KGAE quasibiennial mode. Niño3.4 references the Niño3.4 Index, PDO references the Pacific Decadal Oscillation index, E- and C- Indices follow \citet{takahashi_e_and_c_patterns} and reference Eastern-Pacific and Central-Pacific ENSO Flavors respectively, PMM references the Pacific Meridional Mode index of \citet{chiang_analogous_2004}, and NPGO references the North Pacific Gyre Oscillation index of \citet{di_lorenzo_north_2008}.}\label{tab:Atable}
\end{table}

\begin{table}[t]
\centering
\small
\setlength{\tabcolsep}{4pt}
\renewcommand{\arraystretch}{1.15}
\textbf{Panel A: Full basin}\\
\begin{tabular}{lcccccc}
\toprule
mode & C-Index & E-Index & NPGO & ONI & PDO & PMM \\
kgae &  &  &  &  &  &  \\
\midrule
$DEC_{IA}$ & -0.51 [-0.68, -0.23] & +0.15 [-0.12, +0.37] & +0.76 [+0.54, +0.79] & -0.34 [-0.53, -0.10] & -0.39 [-0.57, -0.14] & \textbf{-0.86 [-0.84, -0.69]} \\
$DEC_{lin}$ & +0.72 [+0.68, +0.75] & +0.23 [+0.17, +0.30] & -0.77 [-0.79, -0.74] & +0.61 [+0.57, +0.66] & \textbf{+0.84 [+0.80, +0.87]} & +0.78 [+0.74, +0.82] \\
$DEC_{self}$ & -0.44 [-0.55, -0.30] & +0.09 [-0.08, +0.23] & +0.63 [+0.51, +0.70] & -0.28 [-0.40, -0.15] & -0.61 [-0.71, -0.48] & \textbf{-0.84 [-0.85, -0.76]} \\
$IA_{DEC}$ & +0.04 [-0.18, +0.22] & -0.48 [-0.59, -0.30] & -0.38 [-0.52, -0.16] & -0.14 [-0.33, +0.04] & +0.15 [-0.05, +0.30] & \textbf{+0.77 [+0.60, +0.79]} \\
$IA_{lin}$ & +0.81 [+0.77, +0.84] & +0.80 [+0.76, +0.82] & -0.43 [-0.48, -0.37] & \textbf{+0.81 [+0.78, +0.83]} & +0.60 [+0.55, +0.64] & -0.12 [-0.17, -0.05] \\
$IA_{self}$ & -0.58 [-0.68, -0.34] & \textbf{-0.61 [-0.68, -0.43]} & +0.21 [-0.06, +0.41] & -0.58 [-0.66, -0.35] & -0.27 [-0.41, -0.10] & +0.26 [+0.05, +0.48] \\
$QB_{lin}$ & +0.70 [+0.49, +0.80] & \textbf{+0.71 [+0.48, +0.79]} & -0.55 [-0.62, -0.35] & +0.52 [+0.24, +0.71] & +0.52 [+0.31, +0.61] & -0.01 [-0.10, +0.10] \\
\bottomrule
\end{tabular}
\vspace{0.6em}
\textbf{Panel B: Tropics (23S-23N)}\\
\begin{tabular}{lcccccc}
\toprule
mode & C-Index & E-Index & NPGO & ONI & PDO & PMM \\
kgae &  &  &  &  &  &  \\
\midrule
$DEC_{IA}$ & -0.46 [-0.66, -0.17] & +0.31 [+0.01, +0.53] & +0.71 [+0.45, +0.82] & -0.30 [-0.53, -0.03] & -0.39 [-0.59, -0.13] & \textbf{-0.91 [-0.90, -0.71]} \\
$DEC_{lin}$ & +0.69 [+0.63, +0.74] & +0.05 [-0.04, +0.15] & \textbf{-0.88 [-0.90, -0.84]} & +0.53 [+0.47, +0.60] & +0.74 [+0.68, +0.80] & +0.76 [+0.69, +0.81] \\
$DEC_{self}$ & -0.32 [-0.50, -0.12] & +0.36 [+0.13, +0.52] & +0.59 [+0.42, +0.73] & -0.12 [-0.29, +0.05] & -0.34 [-0.52, -0.15] & \textbf{-0.89 [-0.89, -0.77]} \\
$IA_{DEC}$ & -0.08 [-0.32, +0.17] & -0.69 [-0.77, -0.49] & -0.21 [-0.41, +0.03] & -0.26 [-0.45, -0.04] & -0.12 [-0.33, +0.11] & \textbf{+0.84 [+0.66, +0.85]} \\
$IA_{lin}$ & \textbf{+0.84 [+0.80, +0.87]} & +0.83 [+0.79, +0.85] & -0.59 [-0.64, -0.55] & +0.84 [+0.81, +0.87] & +0.79 [+0.76, +0.82] & -0.26 [-0.31, -0.17] \\
$IA_{self}$ & -0.72 [-0.80, -0.47] & -0.73 [-0.76, -0.55] & +0.47 [+0.16, +0.62] & \textbf{-0.76 [-0.80, -0.50]} & -0.62 [-0.72, -0.39] & +0.30 [+0.04, +0.52] \\
$QB_{lin}$ & +0.77 [+0.65, +0.84] & \textbf{+0.79 [+0.63, +0.84]} & -0.56 [-0.68, -0.43] & +0.63 [+0.44, +0.78] & +0.70 [+0.58, +0.75] & -0.26 [-0.36, -0.08] \\
\bottomrule
\end{tabular}
\vspace{0.6em}
\textbf{Panel C: North Pacific (21N+)}\\
\begin{tabular}{lcccccc}
\toprule
mode & C-Index & E-Index & NPGO & ONI & PDO & PMM \\
kgae &  &  &  &  &  &  \\
\midrule
$DEC_{IA}$ & -0.75 [-0.80, -0.52] & -0.25 [-0.45, -0.01] & \textbf{+0.84 [+0.65, +0.83]} & -0.35 [-0.52, -0.13] & -0.48 [-0.70, -0.19] & -0.81 [-0.83, -0.59] \\
$DEC_{lin}$ & +0.88 [+0.87, +0.89] & +0.63 [+0.61, +0.65] & -0.65 [-0.68, -0.61] & +0.64 [+0.62, +0.69] & \textbf{+0.92 [+0.91, +0.94]} & +0.87 [+0.85, +0.88] \\
$DEC_{self}$ & -0.76 [-0.79, -0.69] & -0.52 [-0.57, -0.44] & +0.66 [+0.57, +0.71] & -0.42 [-0.54, -0.34] & \textbf{-0.84 [-0.88, -0.76]} & -0.82 [-0.84, -0.77] \\
$IA_{DEC}$ & +0.55 [+0.37, +0.64] & +0.27 [+0.11, +0.37] & -0.68 [-0.73, -0.51] & +0.08 [-0.08, +0.22] & +0.55 [+0.35, +0.67] & \textbf{+0.70 [+0.55, +0.75]} \\
$IA_{lin}$ & +0.29 [+0.14, +0.42] & +0.44 [+0.31, +0.54] & +0.22 [+0.06, +0.33] & \textbf{+0.70 [+0.59, +0.78]} & +0.25 [+0.08, +0.42] & -0.07 [-0.21, +0.06] \\
$IA_{self}$ & +0.38 [+0.09, +0.52] & +0.20 [-0.06, +0.34] & -0.50 [-0.66, -0.15] & +0.01 [-0.25, +0.21] & +0.50 [+0.13, +0.64] & \textbf{+0.55 [+0.20, +0.65]} \\
$QB_{lin}$ & +0.50 [+0.23, +0.59] & +0.47 [+0.05, +0.64] & \textbf{-0.53 [-0.68, -0.15]} & +0.19 [-0.17, +0.45] & +0.47 [-0.00, +0.65] & +0.42 [+0.14, +0.51] \\
\bottomrule
\end{tabular}
\vspace{0.6em}
\caption{Spatial correlations between KGAE patterns and known climate-mode regression patterns. Entries show median correlation across seeds with 5--95\% interval in brackets.}
\label{tab:kgae_climate_panels}
\end{table}
\clearpage

\bibliographystyle{ametsocV6}
\bibliography{references_r2}

\includepdf[pages=-]{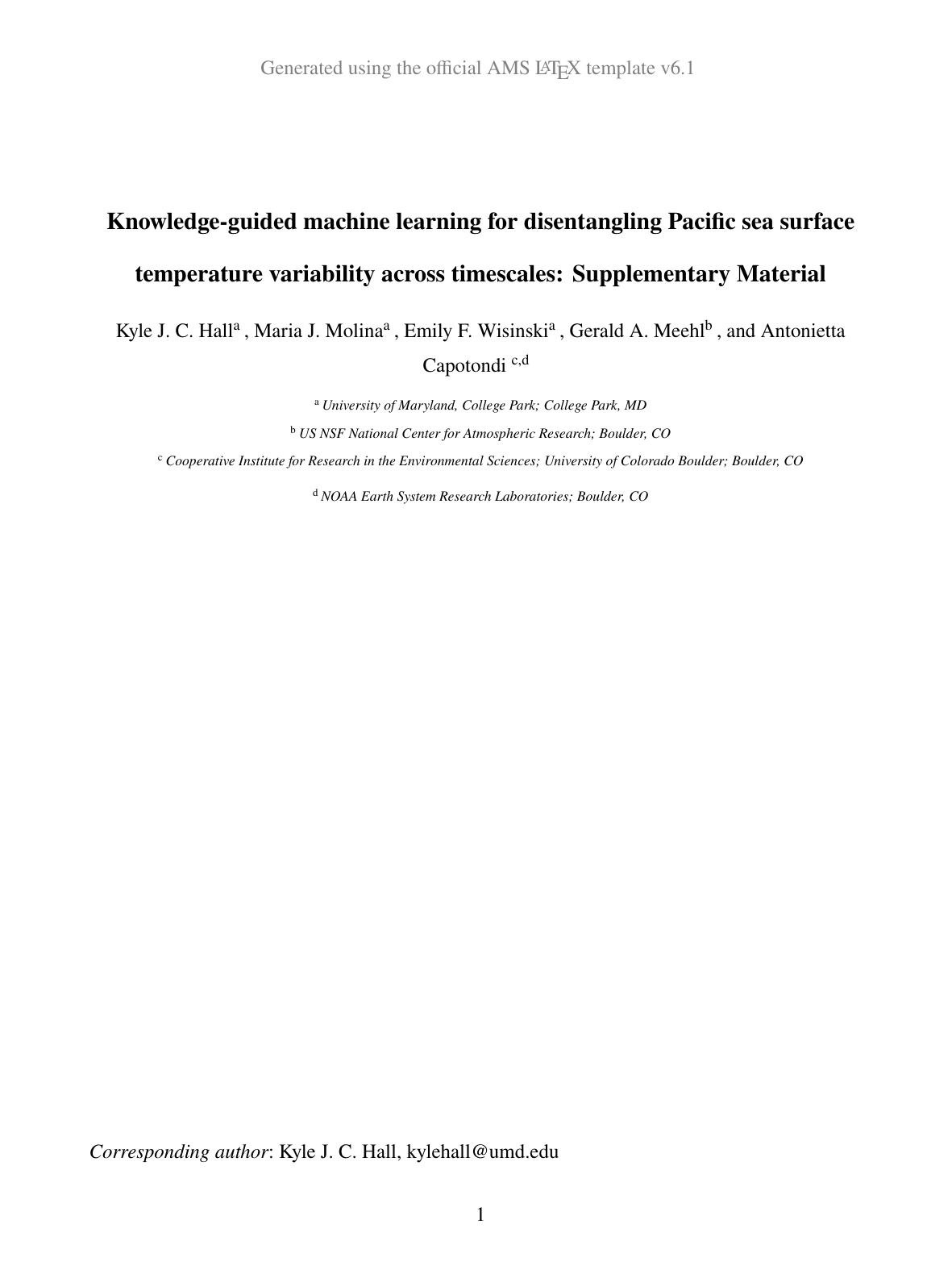}
\end{document}